\documentclass[a4paper,11pt]{article}
\usepackage[left=2cm,right=2cm,top=2cm,bottom=2cm]{geometry}
\usepackage[usenames,dvipsnames,svgnames,table]{xcolor}
\usepackage{psfrag,amsmath,amsfonts,amssymb,latexsym,amsthm,verbatim,color,lscape,multirow,graphicx,subcaption}
\usepackage[figuresright]{rotating}
\usepackage[compress]{natbib}
\usepackage{url}
\usepackage{hyperref}
\hypersetup{colorlinks,
citecolor=black,
filecolor=black,
linkcolor=black,
urlcolor=black,
pdftex}
\usepackage{pdfpages}
\usepackage{booktabs}
\usepackage{tikzducks}

\usepackage{colortbl}
\definecolor{gray}{rgb}{0.9,0.9,0.9}

\usepackage{times}
\usepackage{blindtext}
\usepackage[font={footnotesize,it}]{caption}

\usepackage{sectsty}  
\subsectionfont{\normalsize\bf}
\sectionfont{\large\bf}

\usepackage[symbol]{footmisc}

\usepackage{algorithm}
\usepackage[noend]{algpseudocode}

\def\bfpi{\boldsymbol{\pi}}

\def\rt#1{\sqrt{#1}\,}

\def\R{\mathbf{R}}

\def\s{\sigma}

\def\a{{\alpha}}
\def\g{\gamma}
\def\b{\beta}

\def\de{\delta}
\def\debf{\boldsymbol{\delta}}

\def\th{\theta}

\def\thbf{\boldsymbol{\theta}}
\def\taubf{\boldsymbol{\tau}}

\def\Sbf{\boldsymbol{\Sigma}}

\def\bmu{\boldsymbol{\mu}}

\begin{document}

\title{An one-factor copula mixed model for joint meta-analysis \\of multiple diagnostic tests}

\author{Aristidis~K.~Nikoloulopoulos \footnote{\href{mailto:a.nikoloulopoulos@uea.ac.uk}{a.nikoloulopoulos@uea.ac.uk},  School of Computing Sciences, University of East Anglia, Norwich NR4 7TJ, U.K.} }
\date{}

\maketitle

\begin{abstract}
\baselineskip=24pt
\noindent As the meta-analysis of  more than one diagnostic tests can impact clinical decision making and patient health, there is an increasing body of research in models and methods for meta-analysis of  studies  comparing multiple diagnostic tests. 
The application of the existing  models  to compare  the accuracy of three or more tests  suffers from the curse of multi-dimensionality, i.e.,  either  the number of model parameters increase rapidly or high dimensional integration is required. To overcome these  issues in joint meta-analysis of  studies comparing  $T >2$ diagnostic tests 
in a multiple tests design with a gold standard, we propose a model that assumes the true positives and true negatives for each test are conditionally independent and binomially distributed given the  $2T$-variate latent  vector of  sensitivities and specificities. For the random effects distribution, we employ an one-factor copula that  provides  
  tail dependence or tail asymmetry. Maximum likelihood estimation of the model is straightforward  as the derivation of the likelihood requires bi-dimensional  instead of  $2T$-dimensional integration. 
Our  methodology is demonstrated with an extensive simulation study and an application example  that determines which is the best test for the diagnosis of rheumatoid arthritis.

\noindent \textbf{Key Words:} Diagnostic tests; factor copulas; multivariate meta-analysis;   mixed models; sensitivity/specificity, summary receiver operating characteristic curves. 

\end{abstract}

\section{Introduction}
\baselineskip=24pt

The  identification of the most accurate diagnostic test for a particular disease contributes to the prevention of unnecessary risks to patients and  healthcare costs.   Diagnostic test accuracy studies aim to identify a new diagnostic test that is as 
accurate as the current perfect reference standard, also known as gold standard, yet less expensive or invasive.

Clinical and policy decisions are usually made on the basis of the results from many  diagnostic test accuracy studies on  the same research question. 
The considerably large  number of diagnostic test accuracy  studies has led to the use of meta-analysis.
The purpose of a meta-analysis of diagnostic test accuracy studies is to combine information over different studies, and provide an integrated analysis that will have more statistical power to detect an accurate diagnostic test than an analysis based on a single study. As the accuracy of a diagnostic test
is commonly measured by a pair of indices such as sensitivity and specificity, synthesis of diagnostic test accuracy studies is the most common medical application of multivariate meta-analysis 
\citep{JacksonRileyWhite2011}. 
Most of the existing meta-analysis models and methods, when a perfect reference standard is available,
have mainly focused on a single test (e.g., \citealt{RutterGatsonis2001,Reitsma-etal-2005,Chu&Cole2006}).

However,  as the understanding of a particular disease increases, along with  technological advances,  the  comparative test accuracy of more than one  diagnostic tests  is apparent.  
As summarized by \cite{Takwoingi-etal-2013},  diagnostic test accuracy studies can be  comparative when they assess two or more tests or non-comparative when they assess one diagnostic  test. Estimates of comparative test accuracy can be obtained from either category of studies, but the ones from the latter category are confounded by study setting. 
The robust comparative studies of diagnostic test accuracy 
use either  a multiple test (also called paired or crossover)   design, 
 in which all patients undergo all tests
together with the perfect reference standard, or more rarely, a randomised (also called  parallel) design,
in which all patients undergo the perfect reference standard test but are randomly allocated  to have only one of the other tests. 
A multiple test design is statistically much more efficient, in that one needs much smaller sample sizes to detect a given difference in test accuracy, compared with a randomized design.

As the meta-analysis of more than one diagnostic tests can impact clinical decision making and patient health
 there is an increasing body of research that focus on the development  of meta-analysis models and methods for the synthesis of   studies comparing multiple diagnostic tests. 
\cite{trikalinos-etal-2014-rsm} were the  first who developed a model for the joint meta-analysis of studies comparing two diagnostic tests 
in a multiple tests design with a gold standard. 
They proposed a multinomial generalized linear mixed model  (GLMM) which assumes independent multinomial distributions for the counts of each combination of test results in diseased patients, and, the counts of each combination of test results in non-diseased patients, conditional on the transformed latent  true positive rate (TPR) and false positive rate (FPR) for each test, and latent joint TPR and  FPR, which capture information on the agreement between the two tests  in each study. 
\cite{dimou-etal2016}  extended the bivariate model of \cite{Reitsma-etal-2005}, which jointly meta-analyses the study-estimates of sensitivity and specificity for the case of a single test, to the case of two tests. They modelled the transformed study-estimates of   TPR and FPR  of the two tests using a quadrivariate normal distribution, with the information 
on the agreement between the two tests incorporated in the calculation of the within-study covariance matrix which is assumed fixed. 
\cite{Nikoloulopoulos2020-6variateREMADA}  proposed a multinomial truncated D-vine copula mixed model for the joint meta-analysis of studies comparing two diagnostic tests, which assumes independent multinomial distributions for the counts of each combination of test results in diseased and non-diseased patients, conditional on the latent vector of probabilities of each combination of test results in diseased and non-diseased patients. 
Their proposed model includes the multinomial GLMM  \citep{trikalinos-etal-2014-rsm} as a special case, but can also operate on the original scale of the latent proportions.

As the information on the agreement between the two tests is usually not available from all the primary studies,  \cite{hoyer&kuss-2016-smmr} proposed a model that is solely  based on the information  from the two (one per test) $2\times2$ tables with the number of true positives, true negatives, false negatives  and false positives  per study. They extended the bivariate generalized mixed model (GLMM) proposed by  \cite{Chu&Cole2006} to the quadrivariate case.   The proposed quadrivariate GLMM assumes that the true positives and true negatives from the two tests  are conditionally independent and binomially distributed given  the bivariate latent pairs of transformed sensitivity and specificity, which are  quadrivariate normally distributed.  
 \cite{nikoloulopoulos-2018-smmr}  generalised the quadrivariate GLMM by proposing a model that instead links the four random effects using a quadrivariate D-vine copula rather than the quadrivariate normal  distribution.

However,  for a particular disease there may be three (or more) diagnostic tests 
developed, where each of the tests is subject to several studies (e.g., \citealt{Takwoingi-etal-2013}).  
The extension of the aforementioned  models \citep{trikalinos-etal-2014-rsm,
dimou-etal2016,hoyer&kuss-2016-smmr,nikoloulopoulos-2018-smmr,Nikoloulopoulos2020-6variateREMADA} to compare  the accuracy of more than two tests suffers from the curse of multi-dimensionality, i.e.,  either  the number of model parameters increase rapidly or high dimensional integration is required.

In this paper to overcome the  drawbacks  in existing models for the joint meta-analysis of  studies  comparing $T >2$ diagnostic tests 
 in a multiple test  design with a gold standard, we propose a model that assumes the true positives and true negatives for each test are conditionally independent and binomially distributed given the  $2T$-variate latent (random) vector of (transformed) sensitivities and specificities. 
For the random effects distribution, we employ an one-factor copula \citep{krupskii-joe-2013,nikoloulopoulos&joe12,Kadhem&Nikoloulopoulos-2021}.   The one-factor copula can provide, 
with appropriately chosen linking copulas,
 asymmetric dependence structure as well as tail dependence (dependence among extreme values)
as it is an 1-truncated C-vine copula \citep{Brechmann-Czado-Aas-2012}  rooted at the latent variable/factor. \cite{joeetal10} have shown that  by choosing bivariate linking copulas appropriately, vine copulas  can have a flexible range of lower/upper tail dependence, and different lower/upper tail dependence parameters for each bivariate margin. 
With an one-factor copula, dimension reduction is achieved as the dependence among the latent sensitivities and specificities is explained  by one other latent variable/factor. Hence, the proposed model has $2T$ dependence parameters  instead of $T(2T-1)$, but more importantly its derivation requires bi-dimensional  instead of  $2T$-dimensional integration.

The remainder of the paper proceeds as follows. 
Section \ref{copula-mixed-model-sec}  introduces the one-factor copula mixed model for the comparison of multiple diagnostic tests in a multiple tests design with a gold standard and Section
\ref{relation2GLMM}  discusses its relationship with the $2T$-variate GLMM. Section  \ref{SROC-section} deduces summary receiver operating characteristic (SROC) curves   from the proposed model through quantile regression techniques.  Section \ref{computation} provides a fast and efficient  maximum likelihood (ML) estimation technique based on dependent  Gauss-Legendre quadrature points that have an one-factor copula distribution  and  Section \ref{miss-section} contains small-sample efficiency calculations to investigate the effect of misspecifying the random effects distribution on parameter estimators and standard errors.  Section \ref{app-sec} applies   our methodology  to data from a meta-analysis of diagnostic tests for rheumatoid arthritis. We conclude with some discussion in Section \ref{discussion}, followed by a brief section with software details.

\section{\label{copula-mixed-model-sec}The one-factor copula mixed model }

We first introduce the notation used in this paper. Let  $i$  be  an
index for the individual studies, $j$  an index for the test outcome (0:negative; 1:positive), $k$   an index for the disease outcome (0: non-diseased; 1: diseased) and $t$  an index for the diagnostic test. The frequency data  
 $y_{ijkt},\, i = 1, . . . ,N,\, j=0,1,\,k=0,1,\,t=1,\ldots,T$,  corresponding to a combination of index test  and disease outcomes in study $i$  for test $t$, form a $2\times 2T$ table (Table \ref{2times2T}), that is $T$  ``classic" $2\times 2$ tables.  We assume that the gold standard is the same for the $T$ tests, i.e. $y_{i+01}=\cdots=y_{i+0T}$ and $y_{i+11}=\cdots=y_{i+1T}$.

\begin{table}[!h]
\caption{\label{2times2T} Data  from an individual study in a  $2\times 2T$  table. }
\centering
\begin{small}
\setlength{\tabcolsep}{8.5pt}

\begin{tabular}{llccccllccccllcccc}
\toprule
&&\multicolumn{2}{c}{Disease }&&&&\multicolumn{2}{c}{Disease }&&&&\multicolumn{2}{c}{Disease }\\ 
{Test 1}&  & $-$ & $+$&$\ldots$ &{Test $t$}&  & $-$ & $+$&$\ldots$ &{Test $T$}&  & $-$ & $+$\\\hline
$-$&  & $y_{i001}$ &$y_{i011}$ &$\ldots$ &$-$&  & $y_{i00t}$ &$y_{i01t}$ &$\ldots$ &$-$&  & $y_{i00T}$ &$y_{i01T}$  \\
$+$ & & $y_{i101}$&  $y_{i111}$ & $\ldots$ &$+$ & & $y_{i10t}$&  $y_{i11t}$ & $\ldots$ &$+$ & & $y_{i10T}$&  $y_{i11T}$ \\
\hline
Total &  &$y_{i+01}$&$y_{i+11}$&$\ldots$ &Total &  &$y_{i+0t}$&$y_{i+1t}$&$\ldots$ &Total &  &$y_{i+0T}$&$y_{i+1T}$\\
\bottomrule
\end{tabular}
\end{small}

\end{table}

The within-study model assumes that the number of true positives $Y_{i11t}$ and true negatives $Y_{i00t}$ for  $t=1,\ldots,T$  are conditionally independent and binomially distributed given $(X_{1t},X_{0t})=(x_{1t},x_{0t})$, where $(X_{1t},X_{0t})$ denotes the  bivariate latent  pair of  (transformed)  sensitivity and specificity for the  test $t$.  
That is
\begin{eqnarray}\label{withinBinom}
Y_{i11t}|X_{1t}=x_{1t}&\sim& \mbox{Binomial}\Bigl(y_{i+1t},l^{-1}(x_{1t})\Bigr);\nonumber\\
Y_{i00t}|X_{0t}=x_{0t}&\sim& \mbox{Binomial}\Bigl(y_{i+0t},l^{-1}(x_{0t})\Bigr),
\end{eqnarray}
for $t=1,\ldots, T$, where $l(\cdot)$ is a link function.

For the between studies model,
there are different latent variables $(X_{1t},X_{0t})$
for each test, but they are dependent. Hence the observed data  $y_{ijkt}$ are dependent.
In multivariate models with copulas, a copula or multivariate uniform distribution is combined with a set of univariate margins \citep{joe2014}. This is equivalent to assuming that the latent variables $X_{kt},\,k=0,1,\,t=1,\ldots,T$ have been transformed to standard uniform latent variables $U_{kt}=F\bigl(X_{kt};l(\pi_{kt}),\de_{kt}\bigr),\,k=0,1,\,t=1,\ldots,T$, where $F\bigl(\cdot;l(\pi),\de\bigr)$ is the cumulative distribution function (cdf) of the univariate distribution of the random effect. So we assume that $(U_{11},\ldots,U_{1T},U_{01},\ldots,U_{0T})$ is a random vector with $U_{kt}\sim U(0,1)$. The joint cdf is then given by $C (u_{11},\ldots,u_{1T},u_{01},\ldots,u_{0T})$ where $C$ is a $2T$-dimensional copula. As the joint distribution in general
involves $2T$-dimensional numerical integration, we avoid multidimensional integration 
via an 1-factor copula model whose joint distribution involves only 1-dimensional integration.
In the one-factor copula model, $U_{11},\ldots,U_{1T},U_{01},\ldots,U_{0T}$ are assumed to be conditionally independent given another  latent variable $V$ that is also standard uniformly distributed. For $k=0,1,\,t=1,\ldots,T$ denote the joint cdf and density of $(U_{kt},V)$ by $C_{kt,V}(u_{kt},v;\th_{kt})$ and $c_{kt,V}(u_{kt},v;\th_{kt})=\frac{\partial C_{kt,V}(u_{kt},v;\th_{kt})}{\partial u_{kt}\partial v}$, respectively, and the conditional copula  cdf of $U_{kt}|V$ by $C_{kt|V}(u_{kt}|v;\th_{kt})=\frac{\partial C_{kt,V}(u_{kt},v;\th_{kt})}{\partial v}$. The parameters $\th_{kt}$ are the bivariate copula parameters and are separated from the marginal parameters $\pi_{kt},\de_{kt}$. 
Then, the $2T$-dimensional  one-factor  copula cdf and density with dependence parameter vector $\thbf=(\th_{11},\ldots,\th_{1t},\ldots,\th_{1T},\th_{01},\ldots,\th_{0t},\ldots,\th_{0T})$ are 
\begin{equation}\label{1F-cdf}
C(u_{11},\ldots,u_{1t},\ldots,u_{1T},u_{01},\ldots,u_{0t},\ldots,u_{0T};\thbf)=\int_0^1\prod_{t=1}^TC_{1t|V}(u_{1t}|v;\th_{1t})C_{0t|V}(u_{0t}|v;\th_{0t})dv,
\end{equation}
and 
\begin{equation}\label{1F-density}
c(u_{11},\ldots,u_{1t},\ldots,u_{1T},u_{01},\ldots,u_{0t},\ldots,u_{0T};\thbf)=\int_0^1\prod_{t=1}^Tc_{1t,V}(u_{1t},u;\th_{1t})c_{0t,V}(u_{0t},u;\th_{0t})dv,
\end{equation}
respectively \citep{krupskii-joe-2013}.  It is seen that the $2T$-variate density/cdf decomposes in an one-dimensional  integral of a product of $2T$ bivariate copula densities/cdfs.

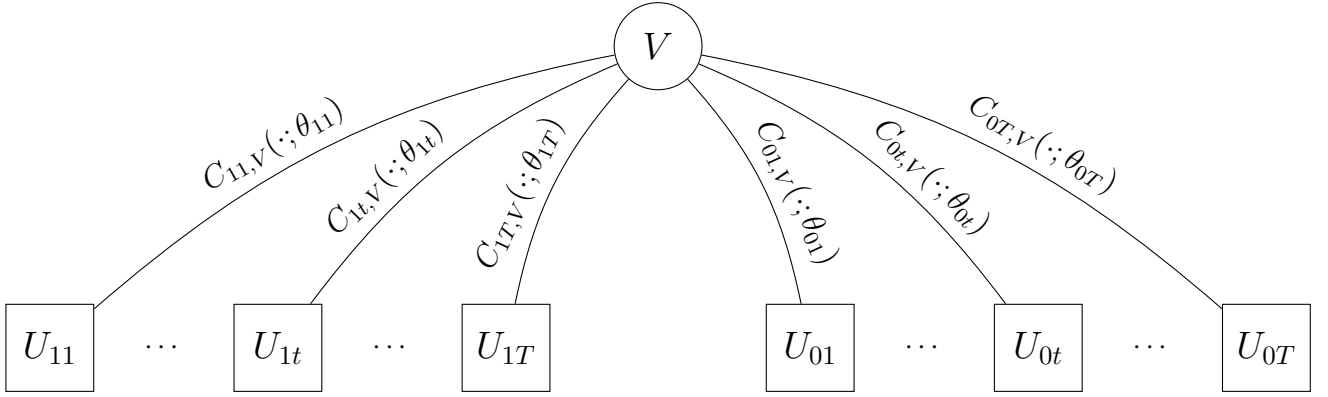
\begin{figure}[h!]
\scalebox{1}{
    \begin{tikzpicture}
        [square/.style={
            draw,
            fill=white!,
            minimum width=3em,
            minimum height=3em,
            node contents={#1}}
            ]

        \node at (0,1) [circle,draw,fill=white!,minimum width=3em] (X0) {\Large $V$};
           		       
	    \node at (-8,-3) (Y11) [square={\Large $U_{11}$}]; 
         		\node at  (-6.5,-3)  {$\cdots$}; 
        \node at (-5,-3) (Yj1) [square={\Large $U_{1t}$}]; 
                  \node at  (-3.5,-3)  {$\cdots$}; 
        \node at (-2,-3) (Yd11) [square={\Large $U_{1T}$}]; 
	    
\path(X0)edge[black,bend right=15] node[above,rotate=30, midway,pos=0.60]{\large $C_{11,V}(
\cdot;\theta_{11})$}  (Y11);
\path(X0)edge[black,bend right=15] node[above,rotate=45, midway,pos=0.65]{\large $C_{1t,V}(
\cdot;\theta_{1t})$} (Yj1);
\path(X0)edge[black,bend right=15] node[above,rotate=65, midway,pos=0.60]{\large $C_{1T,V}(
\cdot;\theta_{1T})$} (Yd11);
        \node at (2,-3) (Y1G) [square={\Large $U_{01}$}];
         		\node at  (3.5,-3)  {$\cdots$};
        \node at (5,-3) (YjG) [square={\Large $U_{0t}$}];
                  \node at  (6.5,-3)  {$\cdots$};
        \node at (8,-3) (YdGG) [square={\Large $U_{0T}$}];
	    
\path(X0) edge[black,bend right=-15] node[above,rotate=-65, midway,pos= 0.60]{\large $C_{01,V}(
\cdot;\theta_{01})$} (Y1G);
\path(X0) edge[black,bend right=-15] node[above,rotate=-45, midway,pos=0.65]{\large $C_{0t,V}(
\cdot;\theta_{0t})$} (YjG);
\path(X0) edge[black,bend right=-15] node[above,rotate=-30, midway,pos=0.60]{\large $C_{0T,V}(
\cdot;\theta_{0T})$} (YdGG);

    \end{tikzpicture}
}
\caption{\label{graphical-representation}Graphical representation of the 1-factor copula model.}
\end{figure}

The one-factor copula can be explained as an 1-truncated C-vine rooted at the latent variable $V$ \citep{krupskii-joe-2013,nikoloulopoulos&joe12,
Kadhem&Nikoloulopoulos-2021}.
$2T$-dimensional C-vine copulas
can cover flexible dependence structures through the specification
of $2T$ bivariate marginal copulas at level 1 and $T(2T-1)$
bivariate conditional copulas at higher levels \citep{nikoloulopoulos&joe&li11}.
For the $2T$-dimensional one-factor copula, the pairs at level 1 are $U,U_{kt}$, for
$k=0,1,\,t=1,\ldots,T$, 
and for  higher levels  the (conditional)
copula pairs are   set to independence. 
That is 
the 1-factor copula has $2T$ bivariate copulas $C_{kt,V}(\cdot;\th_{kt})$ that link $U_{kt},\,k=0,1,\,t=1,\ldots,T$ with $V$  in the 1st level of the vine and independence copulas in all the remaining levels of the vine (truncated after the 1st level). Figure \ref{graphical-representation} depicts the graphical representation of the 1-factor copula model. 
\cite{joeetal10} have shown that in order for a vine copula to have (tail) dependence for all bivariate margins, it is only necessary for the bivariate copulas in level 1 to have (tail) dependence and it is not necessary for the conditional bivariate copulas in levels $2,\ldots,2T$   to have (tail) dependence. Hence, the (tail) dependence between the factor and each of the latent sensitivities/specificities is inherited to the (tail) dependence between the latent sensitivities and specificities.

The stochastic representation of the between studies model takes the form
\begin{eqnarray}\label{copula-between-norm}
&&\Bigl(F\bigl(X_{11};l(\pi_{11}),\de_{11}\bigr),\ldots,
F\bigl(X_{1t};l(\pi_{1t}),\de_{1t}\bigr),\ldots,F\bigl(X_{1T};l(\pi_{1T}),\de_{1T}\bigr)\nonumber\\
&&F\bigl(X_{01};l(\pi_{01}),\de_{01}\bigr),\ldots,
F\bigl(X_{0t};l(\pi_{0t}),\de_{0t}\bigr),\ldots,F\bigl(X_{0T};l(\pi_{0T}),\de_{0T}\bigr)\Bigr)\sim C(\cdot;\thbf). 
\end{eqnarray} 
The parameters  $(\pi_{11},\ldots,\pi_{1t},\ldots,\pi_{1T})$ $:=\bfpi_1$ and $(\pi_{01},\ldots,\pi_{0t},\ldots,\pi_{0T}):=\bfpi_0$ are those of main interest denoting the meta-analytic parameters for the sensitivities and specificities, respectively, while the univariate parameters $(\de_{11},$ $\ldots,\de_{1t},\ldots,\de_{1T}):=\debf_1$ and $(\de_{01},\ldots,\de_{0t},$ $\ldots,\de_{0T}):=\debf_0$ are of secondary interest denoting the between-study variabilities  for the sensitivities and specificities, respectively.
The copula parameter vector $\thbf$ has parameters of the random effects model and they are separated from the univariate parameters $\bfpi_1,\bfpi_0,\debf_1,\debf_0$. 

The models in (\ref{withinBinom}) and (\ref{copula-between-norm}) together specify an one-factor copula mixed  model with  joint likelihood 
\begin{align}\label{mixed-cop-likelihood}
&L(\bfpi_1,\bfpi_0,\debf_1,\debf_0,\thbf)\nonumber\\
&
=\prod_{i=1}^N\int_{[0,1]^{2T}}\biggl\{\prod_{t=1}^T\biggl[ g\biggl(y_{i11t};y_{i+1t},l^{-1}\Bigl(F^{-1}\bigl(u_{1t};l(\pi_{1t}),\de_{1t}\bigr)\Bigr)\biggr) g\biggl(y_{i00t};y_{i+0t},
l^{-1}\Bigl(F^{-1}\bigl(u_{0t};\nonumber\\&l(\pi_{0t}),\de_{0t}\bigr)\Bigr)\biggr)\biggr]
\int_0^1\biggl\{\prod_{t=1}^T\biggl[c_{1t,V}(u_{1t},v;\th_{1t})c_{0t,V}(u_{0t},v;\th_{0t})\biggr]\biggr\}dv\biggr\}du_{11},\ldots,du_{1T}\,du_{01},\ldots,du_{0T}\nonumber\\
&=
\prod_{i=1}^N\int_0^1\biggl\{\prod_{t=1}^T\biggl[ \int_0^1 \biggl\{g\biggl(y_{i11t};y_{i+1t},l^{-1}\Bigl(F^{-1}\bigl(u_{1t};l(\pi_{1t}),\de_{1t}\bigl)\Bigr)\biggr)c_{1t,V}(u_{1t},v;\th_{1t})\biggr\}du_{1t}\nonumber\\
&\int_0^1\biggl\{g\biggl(y_{i00t};y_{i+0t},l^{-1}\Bigl(F^{-1}\bigl(u_{0t};l(\pi_{0t}),\de_{0t}\bigl)\Bigr)\biggr)c_{0t,V}(u_{0t},v;\th_{0t})\biggr\}du_{0t}\biggr]\biggr\}dv
\end{align}
where 
$g\bigl(y;n,\pi\bigr)=\binom{n}{y}\pi^y(1-\pi)^{n-y},\quad y=0,1,\ldots,n,\quad 0<\pi<1,$
 is the binomial probability mass function (pmf). It is shown that the joint likelihood is represented as an one-dimensional integral of a function which in turn is a product of $2T$  one-dimensional integrals. As a result, $2T$-dimensional numerical integration has been avoided.

Our general statistical model allows for selection of bivariate copulas and univariate margins independently, i.e., there are no constraints in the choices of parametric bivariate copulas and univariate margins. In line with our previous contributions in copula mixed models \citep{Nikoloulopoulos2015b,Nikoloulopoulos2015c,Nikoloulopoulos-2016-SMMR,Nikoloulopoulos2018-AStA,nikoloulopoulos-2018-smmr,Nikoloulopoulos-2018-3dmeta-NE,Nikoloulopoulos-2018-4dmeta-NE,Nikoloulopoulos2020-6variateREMADA} we use 
\begin{itemize}
\itemsep=10pt
\item  bivariate parametric copulas with different tail dependence behaviour, namely the bivariate normal  (BVN) with intermediate tail dependence, Frank with tail independence, and Clayton with positive lower tail dependence. For the latter we also use its rotated versions to provide negative upper-lower tail dependence (Clayton rotated by 90$^\circ$), positive upper tail dependence (Clayton rotated by 180$^\circ$) and negative lower-upper tail dependence (Clayton rotated by 270$^\circ$);

\item the choices of  $F\bigl(\cdot;l(\pi),\de\bigr)$ and  $l$ that are given in Table \ref{choices}. With a beta distribution we work on the original scale of   the latent sensitivities and specificities.  
\end{itemize}

\begin{table}[!h]
\begin{center}
\caption{\label{choices}The choices of the  $F\bigl(\cdot;l(\pi),\de\bigr)$ and  $l$ in the one-factor copula mixed model.}
\begin{small}
\setlength{\tabcolsep}{45pt}
\begin{tabular}{cccc}
\toprule $F\bigl(\cdot;l(\pi),\de\bigr)$ & $l$ & $\pi$ & $\de$\\\hline
$N(\mu,\s)$ & logit& $l^{-1}(\mu)$&$\s$\\
Beta$(\pi,\gamma)$ & identity & $\pi$ & $\gamma$\\
\bottomrule
\end{tabular}
\end{small}

\end{center}
\end{table}

\section{\label{relation2GLMM}Relationship with the $2T$-variate GLMM}

We show what happens when all the bivariate copulas $C_{kt,V}(;\th_{kt})$ are BVN   and the univariate distribution of the random effects is the  $N(\mu,\s)$ distribution. 

One can easily deduce that the within-study model in (\ref{withinBinom}) is the same as in the $2T$-variate GLMM. Furthermore, 
when $C_{kt,V}(;\th_{kt})$ are all BVN copulas, then (\ref{1F-cdf}) becomes the copula of the multivariate normal distribution with an one-factor correlation structure. 
Let $C_{kt,V}(;\th_{kt})$ be the BVN copula with  correlation parameter $\th_{kt}$. Let $\Phi$ and $\phi$ denote the standard normal cdf and density function, and let $\Phi_2(\cdot;\rho)$ be the BVN cdf with correlation $\rho$. Then $C_{kt,V}(u,v)=\Phi_2(\Phi^{-1}(u),\Phi^{-1}(v);\th_{kt})$  and $C_{kt|V}(u|v)
  = \Phi\left({\Phi^{-1}(u)-\th_{kt}\Phi^{-1}(v) \over \rt{1-\th_{kt}^2}}\right)
$. For (\ref{1F-cdf}), let $u_{kt}=\Phi(z_{kt})$, where $z_{kt}=\frac{x_{kt}-l(\pi_{kt})}{\s_{kt}}$, to get a $2T$-variate distribution with $N(0,1)$ margins. 
Then
\begin{multline*}
C\Bigl(\Phi(z_{11}),\ldots,\Phi(z_{1t}),
\ldots,\Phi(z_{1T}),
\Phi(z_{01}),\ldots,\Phi(z_{0t}),
\ldots,\Phi(z_{0T});\thbf\Bigr)=\\\int_{0}^1\prod_{t=1}^T \left\{\Phi\left({z_{1t}-\th_{1t}\Phi^{-1}(v) \over \rt{1-\th_{1t}^2}}\right)\Phi\left({z_{0t}-\th_{0t}\Phi^{-1}(v) \over \rt{1-\th_{0t}^2}}\right)\right\}dv
\end{multline*}

\begin{multline}\label{MVN-cdf}
\mbox{or} \quad C\Bigl(\Phi(z_{11}),\ldots,\Phi(z_{1t}),
\ldots,\Phi(z_{1T}),
\Phi(z_{01}),\ldots,\Phi(z_{0t}),
\ldots,\Phi(z_{0T});\thbf\Bigr)=\\
\int_{-\infty}^\infty\prod_{t=1}^T \left\{\Phi\left({z_{1t}-\th_{1t}w \over \rt{1-\th_{1t}^2}}\right)\Phi\left({z_{0t}-\th_{0t}w \over \rt{1-\th_{0t}^2}}\right)\right\}
\phi(w)dw.
\end{multline}
This model is the same as the $2T$-variate normal model with an one-factor correlation  structure 

$$\R=\begin{pmatrix}
1&\cdots&\rho_{11,1T}&\rho_{11,01}&\cdots&\rho_{11,0T}\\
\vdots &\ddots&\vdots& \vdots&\vdots&\vdots\\
\rho_{1T,11}&\cdots&1&\rho_{1T,01}&\cdots&\rho_{1T,0T}\\
\rho_{01,11}&\cdots&\rho_{01,1T}&1&\cdots&\rho_{1T,0T}\\
\vdots &\vdots&\vdots& \vdots&\ddots&\vdots\\
\rho_{0T,11}&\cdots&\rho_{0T,1T}&\rho_{0T,01}&\cdots&1\\
\end{pmatrix}$$

with 
\begin{equation}\label{BVN-rho}
\rho_{k_1t_1,k_2t_2}=\theta_{k_1t_1}\theta_{k_2t_2}, \quad k_1,k_2=0,1,\,t_1,t_2=1,\ldots,T. 
\end{equation} 
This occurs because the multivariate cdf in (\ref{MVN-cdf}) comes from the representation  
\begin{equation}\label{additive-representation}
Z_{kt}=\frac{X_{kt}-l(\pi_{kt})}{\s_{kt}}=\th_{kt}W+\sqrt{1-\th_{kt}^2} \epsilon_{kt},\quad k=0,1,\,t=1,\ldots,T,
\end{equation} 
where $W,\epsilon_{kt}$ are i.i.d. $N(0,1)$ random variables \citep{krupskii-joe-2013,nikoloulopoulos&joe12}.

The resulting random effects distribution for $(X_{11},\ldots,X_{1t},\ldots,X_{1T},X_{01},\ldots,X_{0t},\ldots,X_{0T})$ is the $2T$-variate normal distribution  with mean vector $\bmu=\bigl(l(\bfpi_1),l(\bfpi_0)\bigr)$ and variance-covariance matrix
$$\Sbf=\begin{pmatrix}
\s_{11}^2&\cdots&\rho_{11}\rho_{1T}\s_{11}\s_{1T}&\rho_{11}\rho_{01}\s_{11}\s_{01}&\cdots&\rho_{11}\rho_{0T}\s_{11}\s_{0T}\\
\vdots &\ddots&\vdots& \vdots&\vdots&\vdots\\
\rho_{1T}\rho_{11}\s_{1T}\s_{11}&\cdots&\s_{1T}^2&\rho_{1T}\rho_{01}\s_{1T}\s_{01}&\cdots&\rho_{1T}\rho_{0T}\s_{1T}\s_{0T}\\
\rho_{01}\rho_{11}\s_{01}\s_{11}&\cdots&\rho_{01}\rho_{1T}\s_{01}\s_{1T}&\s_{01}^2&\cdots&\rho_{1T}\rho_{0T}\s_{1T}\s_{0T}\\
\vdots &\vdots&\vdots& \vdots&\ddots&\vdots\\
\rho_{0T}\rho_{11}\s_{0T}\s_{11}&\cdots&\rho_{0T}\rho_{1T}\s_{0T}\s_{1T}&\rho_{0T}\rho_{01}\s_{0T}\s_{01}&\cdots&\s_{0T}^2\\
\end{pmatrix}$$
Hence,  the proposed model has as special case  the $2T$-variate GLMM with an one-factor correlation structure that has a latent additive structure as seen in (\ref{additive-representation}). Nevertheless, if other bivariate copulas are used, then  the one-factor copula mixed model has  a latent structure that is non-additive.

\section{Summary receiver operating characteristic curves}
\label{SROC-section}

Though typically the focus of meta-analysis has been to derive the  summary-effect estimates, there is increasing interest in alternative summary outputs, such as summary receiver operating characteristic (SROC) curves (e.g., \citealt{Arends-etal-2008,Rucker-schumacher-2009}).  
In this section we derive the SROC curves  from the one-factor copula mixed model. 

For the one-factor copula mixed model, the model parameters (including dependence parameters), the choice of the copula, and the choice of the margin will affect the shape of the SROC curve. 
Let the joint cdf of $(U_{1t},U_{0t})$ be given by the copula  $C_{1t,0t}(\cdot;\th_{1t,0t})$.  
The copula parameters $\th_{1t,0t},\,t=1,\ldots,T$ can be derived using the following steps:

\begin{enumerate}
\item Convert the copula parameters $\th_{1t}$ and $\th_{0t}$ of BVN, Frank or   (rotated)  Clayton copulas to Kendall's $\tau_{1t}$ and $\tau_{0t}$  via the relations  
\begin{equation}\label{tauBVN}
\tau=\frac{2}{\pi}\arcsin(\th),
\end{equation}
\begin{equation}\label{tauFrank}
\tau=\left\{\begin{array}{ccc}
1-4\theta^{-1}-4\theta^{-2}\int_\theta^0\frac{t}{e^t-1}dt &,& \th<0\\
1-4\theta^{-1}+4\theta^{-2}\int^\theta_0\frac{t}{e^t-1}dt &,& \th>0\\
\end{array}\right.,
\end{equation}
or
\begin{equation}\label{tauCln}
\tau=\left\{\begin{array}{rcl}
\th/(\th+2)&,& \mbox{by 0$^\circ$ or 180$^\circ$ }\\
-\th/(\th+2)&,& \mbox{by 90$^\circ$ or 270$^\circ$}\\
\end{array}\right.,
\end{equation}
in \cite{HultLindskog02}, \cite{genest87}, or 
\cite{genest&mackay86}, respectively.

\item Convert the Kendall's $\tau_{1t}$ and $\tau_{0t}$ to BVN copula parameters $\th_{1t}$ and $\th_{0t}$ using the inverse of the relation in (\ref{tauBVN}).

\item Convert the BVN  copula parameters $\th_{1t}$ and $\th_{0t}$ 
to the correlation parameter $\rho_{1t,0t}$ 
via the relation in (\ref{BVN-rho}). 

\item Convert the correlation parameter $\rho_{1t,0t}$ to Kendall's $\tau_{1t,0t}$ via the relation (\ref{tauBVN}).

\item  Convert the Kendall's $\tau_{1t,0t}$ to the copula  parameter $\th_{1t,0t}$ of BVN, Frank or   (rotated)  Clayton copula via the inverses of the  relations in (\ref{tauBVN}), (\ref{tauFrank}), or (\ref{tauCln}).
\end{enumerate}

Then, the  SROC
 curves for the  latent pair $(X_{1t}, X_{0t})$  can be deduced 
 through the quantile regression techniques proposed by \cite{Nikoloulopoulos2015b}:
 \begin{enumerate}
\item Set $C_{1t|0t}(u_{1t}|u_{0t};\th_{1t,0t})=q$;
\item Solve for the quantile regression curve $u_{1t}:=\widetilde{u}_{1t}(u_{0t},q;\th_{1t,0t})=C_{1t|0t}^{-1}(q|u_{0t};\th_{1t,0t})$;
\item Replace $u_{kt}$ by $F\Bigl(x_{kt};l(\pi_{kt}),\de_{kt}\Bigr)$; 
\item Plot $x_{1t}:=\widetilde{x}_{1t}(x_{0t},q)$ versus $x_{0t}$. 
\end{enumerate}

As there is no priori reason to regress $X_{1t}$ on $X_{0t}$ instead of the other way around
\citep{Arends-etal-2008},  quantile regression   curves of $X_{0t}$ on $X_{1t}$ are also derived in a similar manner.  We use the median regression curves  ($q=0.5$), along with the quantile regression curves with a focus on high ($q$ = 0.99) and low quantiles ($q$ = 0.01), which are strongly associated with the upper and lower tail dependence, respectively, imposed from each parametric family of bivariate copulas. These can be seen as confidence regions, as per the terminology  in 
\cite{Rucker-schumacher-2009},   of  the median regression curves. 
Finally, in order  to reserve the nature of a bivariate response instead of a univariate response along with a covariate, we plot the corresponding  contour graph of the bivariate copula density. The contour plot can be seen as the predictive region (analogously  to \citealt{Reitsma-etal-2005}) of the estimated pair  $(\pi_{1t},\pi_{0t})$ of the meta-analytic parameters of sensitivity and specificity at test $t$.

\section{\label{computation}Maximum likelihood estimation and computational details}

Estimation of the model parameters $(\bfpi_1,\bfpi_0,\debf_1,\debf_0,\thbf)$   can be approached by the standard maximum likelihood (ML) method, by maximizing the logarithm of the joint likelihood in (\ref{mixed-cop-likelihood}). 
The estimated parameters can be obtained by 
using a quasi-Newton \citep{nash90} method applied to the logarithm of the joint likelihood.  
This numerical  method requires only the objective
function, i.e.,  the logarithm of the joint likelihood, while the gradients
are computed numerically and the Hessian matrix of the second
order derivatives is updated in each iteration. The standard errors (SE) of the ML estimates can be also obtained via the gradients and the Hessian computed numerically during the maximization process.

For one-factor copula mixed models of the form with joint likelihood as in (\ref{mixed-cop-likelihood}), numerical evaluation of the joint pmf can be achieved with the following steps:

\begin{enumerate}
\itemsep=0pt
\item Calculate Gauss-Legendre \citep{Stroud&Secrest1966}  quadrature points $\{u_q: q=1,\ldots,N_q\}$ 
and weights $\{w_q: q=1,\ldots,N_q\}$ in terms of standard uniform.

\item Numerically evaluate the joint pmf
\begin{align*}
&\int_0^1\biggl\{\prod_{t=1}^T\biggl[ \int_0^1 \biggl\{g\biggl(y_{i11t};y_{i+1t},l^{-1}\Bigl(F^{-1}\bigl(u_{1t};l(\pi_{1t}),\de_{1t}\bigl)\Bigr)\biggr)c_{1t,V}(u_{1t},v;\th_{1t})\biggr\}du_{1t}\\
&\int_0^1\biggl\{g\biggl(y_{i00t};y_{i+0t},l^{-1}\Bigl(F^{-1}\bigl(u_{0t};l(\pi_{0t}),\de_{0t}\bigl)\Bigr)\biggr)c_{0t,V}(u_{0t},v;\th_{0t})\biggr\}du_{0t}\biggr]\biggr\}dv 
\end{align*}
in a double sum:

\begin{align*}
&\sum_{q_1=1}^{N_q}\biggl\{w_{q_1}\prod_{t=1}^T\biggl[\sum_{q_2=1}^{N_q} \biggl\{w_{q_2}g\biggl(y_{i11t};y_{i+1t},l^{-1}\Bigl(F^{-1}\bigl(C_{1t|V}^{-1}(u_{q_2}|u_{q_1};\th_{1t});l(\pi_{1t}),\de_{1t}\bigl)\Bigr)\biggr)\biggr\}\\
&\sum_{q_2=1}^{N_q}\biggl\{w_{q_2}g\biggl(y_{i00t};y_{i+0t},l^{-1}\Bigl(F^{-1}\bigl(C_{0t|V}^{-1}(u_{q_2}|u_{q_1};\th_{0t});l(\pi_{0t}),\de_{0t}\bigl)\Bigr)\biggr)\biggr\}\biggr]\biggr\},
\end{align*}
where $C^{-1}_{kt|V}(u|v;\th_{kt})$ is the inverse conditional bivariate copula cdf. Note that  the independent  quadrature points $\{u_{q_1}: q_1=1,\ldots,N_q\}$ and   $\{u_{q_2}: q_2=1,\ldots,N_q\}$ have been converted to dependent quadrature points  that have an one-factor copula distribution $C(\cdot;\thbf)$. 

\end{enumerate}

With Gauss-Legendre quadrature, the same nodes and weights
are used for different functions;
this helps in yielding smooth numerical derivatives for numerical optimization via quasi-Newton \citep{nash90}. Our one-factor copula mixed model for meta-analysis of multiple diagnostic tests   is straightforward computationally as it requires the calculation of a double summation over the quadrature points.

\section{\label{miss-section}Small-sample efficiency -- Misspecification of the random effects distribution}
In this section, we study the small-sample efficiency and robustness of the ML estimation of the one-factor  copula mixed model. In Section \ref{true}, we gauge the small-sample efficiency of the ML method in Section \ref{computation} and investigate the misspecification of either  the parametric margin or bivariate copula of the random effects distribution. In Section \ref{miss}, we investigate the mixed model misspecification by using  the D-vine copula mixed model proposed by \cite{nikoloulopoulos-2018-smmr} as the  true model. That is we include a sensitivity analysis to the conditional independence assumption.

\subsection{\label{true}Misspecification  of the parametric margin or bivariate pair-copulas}

We randomly generated  10,000 samples of size $n = 20, 50, 100$ from an one-factor copula mixed model with both normal and beta margins that jointly  meta-analyses $T=\{2,3,4\}$ diagnostic tests.

The simulation process is as below:

\begin{enumerate}
\itemsep=0pt
 \item Simulate $(u_{11},\ldots,u_{1T},u_{01},\ldots,u_{0T})$ from an one-factor copula $C(;\taubf)$;  $\taubf$ is converted 
to  the copula  parameter vector $\thbf$ of BVN, Frank or   (rotated)  Clayton copulas via the inverses of the relations in (\ref{tauBVN}), (\ref{tauFrank}), or (\ref{tauCln}).

\item For each test $t$ in $1,\ldots,T$ convert  to proportions via 
\begin{eqnarray*}
x_{1t}&=&l^{-1}\biggl(F^{-1}\Bigl(u_{1t};l(\pi_{1t}),\de_{1t}\Bigr)\biggr) \\
x_{0t}&=&l^{-1}\biggl(F^{-1}\Bigl(u_{0t};l(\pi_{0t}),\de_{0t}\Bigr)\biggr)
\end{eqnarray*}
\item Simulate the study size $n$ from a shifted gamma distribution, i.e., $n\sim \mbox{sGamma}(\a=1.2,\b=0.01,\mbox{lag}=30)$ and round off to the nearest integer. 
\item Draw the number of diseased $n_{1}$ from a $B(n,0.4)$ distribution and set  $n_0=n-n_1$ .
\item For each test $t$ in $1,\ldots,T$ generate $y_{11t}$ and $y_{00t}$  from a $B(n_1,x_{1t})$ and  $B(n_0,x_{0t})$ distribution, respectively, and set $y_{01t}=n_1-y_{11t},y_{10t}=n_0-y_{00t}.$
\end{enumerate}

Representative summaries of findings on the performance of the ML method in Section \ref{computation} are given in Tables \ref{sim1} and \ref{sim2} 
for 6-dimensional ($T = 3$) one-factor copula models with normal and beta margins, respectively.
The true (simulated) bivariate copulas are the Clayton  and the Clayton copula rotated by $270^\circ$ to handle the positive and negative dependencies, respectively. 
True sensitivity $\bfpi_1$ and specificity $\bfpi_0$ vectors  are $(0.8,0.7,0.8)$ and $(0.7,0.8,0.7)$,  the variability parameter vectors are $\debf_1=\debf_0=(1,1,1)$ or $\debf_1=\debf_0=(0.1,0.1,0.1)$ for normal or beta margin, respectively, and the Kendall's $\taubf=(0.6,0.7,0.5,-0.3,-0.4,-0.2)$.   Under each margin, 10,000 meta-analysis data sets are simulated with $N=50$ studies in each data set.
We have estimated the one-factor copula mixed model  with different bivariate  copulas and margins. 
Tables \ref{sim1} and \ref{sim2} contain the resultant biases, root mean square errors (RMSEs) and standard deviations (SDs), along with average theoretical variances, scaled by 100, for the MLEs under different copula choices and margins. The theoretical variances of the MLEs are obtained via the gradients and the Hessian that were computed numerically during the maximization process.

Conclusions from the values in the   Tables \ref{sim1} and \ref{sim2}   are the following:

\begin{itemize}
\itemsep=0pt
\item ML   with  the true one-factor copula mixed  model is highly efficient according to the simulated biases,  SDs and RMSEs.

\item The MLEs  of $\bfpi_1,\bfpi_0$ are not robust to margin misspecification, e.g.,  in Table \ref{sim1} (Table \ref{sim2}) where the true univariate margins are normal (beta) the scaled biases for the MLEs of $\pi_{02}$ for the various one-factor copula mixed models with beta (normal) margins range from $-4.16$ ($3.21$) to $-1.86$ (4.70).

\end{itemize}

\setlength{\tabcolsep}{5pt}
\begin{landscape}
\begin{table}[!h]
\caption{\label{sim1}Small sample of sizes $N = 50$ simulations ($10^4$ replications, $n_q=25$) from the one-factor copula mixed model  with normal margins and biases,  root mean square errors (RMSEs) and standard deviations (SDs), along with the square root of the average theoretical variances ($\sqrt{\bar V}$), scaled by 100, for the MLEs  under different copula choices and margins.  The true (simulated) copula distributions are the Clayton  and the Clayton copula rotated by $270^\circ$ to handle the positive and negative dependencies, respectively. }
\begin{small}

    \begin{tabular}{lllcccccccccccccccccc}
    \toprule
   {} & margin & copula & $\pi_{11}$ & $\pi_{12}$ & $\pi_{13}$ & $\pi_{01}$ & $\pi_{02}$ & $\pi_{03}$ & $\s_{11}$ & $\s_{12}$ & $\s_{13}$ & $\s_{01}$ & $\s_{02}$ & $\s_{03}$ & $\tau_{11}$ & $\tau_{12}$ & $\tau_{13}$ & $\tau_{01}$ & $\tau_{02}$ & $\tau_{03}$ \\
    \midrule
  \rowcolor{gray}   Bias  & normal & BVN   & -0.35 & -0.43 & -0.32 & -0.03 & 0.04  & -0.04 & -3.56 & -2.72 & -2.94 & -0.43 & 0.61  & -0.83 & 4.93  & 1.94  & 4.88  & -2.01 & -1.45 & -1.75 \\
          &       & Cln\{$0^\circ,270^\circ$\} & -0.31 & -0.40 & -0.31 & 0.05  & 0.08  & 0.02  & -0.83 & -0.91 & -0.76 & -0.96 & -0.83 & -0.97 & 2.49  & 2.59  & 1.81  & -0.75 & -1.26 & -0.46 \\
   \rowcolor{gray}        &       & Cln\{$180^\circ,90^\circ$\} & -1.38 & -1.98 & -1.12 & 0.30  & 0.50  & 0.08  & 3.64  & 6.30  & 3.37  & 3.70  & 6.85  & 1.46  & 4.27  & 3.31  & -0.07 & 5.52  & 7.56  & 3.65 \\
          &       & Frank & -3.71 & -4.92 & -3.42 & 2.40  & 2.28  & 1.64  & 0.70  & 1.48  & 0.48  & 0.64  & 1.75  & -0.16 & 6.42  & 3.70  & 6.36  & -3.32 & -3.67 & -2.43 \\
  \rowcolor{gray}         & beta  & BVN   & -3.99 & -3.25 & -4.01 & -3.38 & -4.16 & -3.33 & -     & -     & -     & -     & -     & -     & 5.80  & 2.63  & 5.81  & -1.43 & -0.16 & -1.39 \\
          &       & Cln\{$0^\circ,270^\circ$\} & -4.28 & -3.57 & -4.31 & -3.21 & -3.92 & -3.24 & -     & -     & -     & -     & -     & -     & 2.85  & 3.46  & 1.57  & -0.08 & 0.07  & -0.20 \\
 \rowcolor{gray}          &       & Cln\{$180^\circ,90^\circ$\} & -5.42 & -5.02 & -5.21 & -3.28 & -4.14 & -3.30 & -     & -     & -     & -     & -     & -     & 4.59  & 3.71  & 0.59  & 6.71  & 9.89  & 4.73 \\
          &       & Frank & -7.03 & -7.04 & -6.82 & -1.21 & -1.86 & -1.88 & -     & -     & -     & -     & -     & -     & 6.89  & 3.92  & 6.87  & -2.63 & -2.90 & -1.85 \\
  \rowcolor{gray}   SD    & normal & BVN   & 2.51  & 3.25  & 2.50  & 3.13  & 2.45  & 3.14  & 12.50 & 12.18 & 12.55 & 12.09 & 12.79 & 11.84 & 12.28 & 13.01 & 11.35 & 11.24 & 11.30 & 11.54 \\
          &       & Cln\{$0^\circ,270^\circ$\} & 2.55  & 3.31  & 2.55  & 3.15  & 2.47  & 3.16  & 12.30 & 11.77 & 12.35 & 11.79 & 12.21 & 11.75 & 12.46 & 13.93 & 10.67 & 10.39 & 10.78 & 10.34 \\
  \rowcolor{gray}         &       & Cln\{$180^\circ,90^\circ$\} & 2.90  & 3.77  & 2.84  & 3.29  & 2.59  & 3.23  & 16.30 & 16.66 & 15.73 & 14.17 & 15.81 & 13.12 & 21.58 & 22.46 & 17.94 & 12.96 & 13.73 & 13.30 \\
          &       & Frank & 3.58  & 4.49  & 3.52  & 3.65  & 2.81  & 3.53  & 14.00 & 13.67 & 13.91 & 12.54 & 13.28 & 12.13 & 11.93 & 12.00 & 11.71 & 12.02 & 11.98 & 12.19 \\
  \rowcolor{gray}         & beta  & BVN   & 2.46  & 2.86  & 2.46  & 2.77  & 2.44  & 2.79  & 2.82  & 2.91  & 2.84  & 2.83  & 2.85  & 2.79  & 11.16 & 11.62 & 10.56 & 10.52 & 10.24 & 11.06 \\
          &       & Cln\{$0^\circ,270^\circ$\} & 2.54  & 2.95  & 2.56  & 2.81  & 2.48  & 2.82  & 2.91  & 2.85  & 2.91  & 2.71  & 2.65  & 2.75  & 12.13 & 13.44 & 10.50 & 9.67  & 9.89  & 9.88 \\
  \rowcolor{gray}         &       & Cln\{$180^\circ,90^\circ$\} & 2.92  & 3.31  & 2.87  & 2.81  & 2.47  & 2.82  & 3.82  & 4.12  & 3.71  & 3.28  & 3.47  & 3.02  & 20.57 & 21.25 & 16.83 & 12.63 & 13.24 & 12.91 \\
          &       & Frank & 3.47  & 3.88  & 3.43  & 3.10  & 2.63  & 3.06  & 3.35  & 3.40  & 3.32  & 2.86  & 2.83  & 2.82  & 10.82 & 10.45 & 10.89 & 11.41 & 11.19 & 11.86 \\
 \rowcolor{gray}    $\sqrt{\bar V}$  & normal & BVN   & 2.40  & 3.09  & 2.41  & 3.08  & 2.40  & 3.07  & 11.73 & 11.43 & 11.93 & 11.36 & 11.85 & 11.22 & 10.09 & 11.13 & 9.26  & 9.47  & 9.25  & 9.86 \\
          &       & Cln\{$0^\circ,270^\circ$\} & 2.32  & 2.95  & 2.34  & 2.95  & 2.25  & 2.99  & 10.79 & 10.39 & 11.08 & 10.77 & 10.95 & 10.86 & 10.47 & 12.64 & 9.04  & 8.59  & 8.86  & 8.83 \\
  \rowcolor{gray}         &       & Cln\{$180^\circ,90^\circ$\} & 2.66  & 3.45  & 2.64  & 3.18  & 2.48  & 3.14  & 13.57 & 13.52 & 13.50 & 12.68 & 13.58 & 12.18 & 12.37 & 12.28 & 11.66 & 10.17 & 10.23 & 10.75 \\
          &       & Frank & 2.66  & 3.27  & 2.68  & 3.00  & 2.22  & 3.06  & 12.57 & 12.12 & 12.66 & 11.50 & 12.00 & 11.28 & 9.20  & 9.49  & 9.08  & 9.80  & 9.36  & 10.39 \\
   \rowcolor{gray}        & beta  & BVN   & 2.29  & 2.70  & 2.30  & 2.69  & 2.29  & 2.67  & 2.52  & 2.73  & 2.56  & 2.68  & 2.51  & 2.65  & 9.96  & 10.92 & 9.04  & 9.43  & 9.21  & 9.83 \\
          &       & Cln\{$0^\circ,270^\circ$\} & 2.17  & 2.54  & 2.19  & 2.55  & 2.13  & 2.59  & 2.33  & 2.46  & 2.38  & 2.46  & 2.20  & 2.51  & 10.89 & 13.21 & 9.33  & 8.52  & 8.82  & 8.76 \\
   \rowcolor{gray}        &       & Cln\{$180^\circ,90^\circ$\} & 2.56  & 2.99  & 2.55  & 2.74  & 2.36  & 2.72  & 3.04  & 3.32  & 3.03  & 3.06  & 3.01  & 2.91  & 12.24 & 12.41 & 11.36 & 10.36 & 10.44 & 10.91 \\
          &       & Frank & 2.45  & 2.77  & 2.48  & 2.60  & 2.10  & 2.65  & 2.79  & 2.96  & 2.80  & 2.64  & 2.42  & 2.60  & 9.22  & 9.65  & 9.05  & 9.86  & 9.47  & 10.44 \\
    RMSE  & normal & BVN   & 2.54  & 3.28  & 2.52  & 3.13  & 2.45  & 3.14  & 13.00 & 12.48 & 12.89 & 12.10 & 12.80 & 11.87 & 13.23 & 13.16 & 12.36 & 11.42 & 11.39 & 11.67 \\
          &       & Cln\{$0^\circ,270^\circ$\} & 2.57  & 3.33  & 2.57  & 3.15  & 2.47  & 3.16  & 12.33 & 11.81 & 12.37 & 11.83 & 12.24 & 11.79 & 12.70 & 14.17 & 10.82 & 10.42 & 10.85 & 10.35 \\
   \rowcolor{gray}        &       & Cln\{$180^\circ,90^\circ$\} & 3.21  & 4.26  & 3.05  & 3.31  & 2.64  & 3.23  & 16.70 & 17.81 & 16.09 & 14.64 & 17.22 & 13.20 & 22.00 & 22.70 & 17.94 & 14.08 & 15.67 & 13.80 \\
          &       & Frank & 5.15  & 6.66  & 4.91  & 4.37  & 3.62  & 3.89  & 14.02 & 13.75 & 13.91 & 12.56 & 13.39 & 12.13 & 13.55 & 12.56 & 13.32 & 12.47 & 12.53 & 12.43 \\
    \rowcolor{gray}       & beta  & BVN   & 4.69  & 4.33  & 4.70  & 4.37  & 4.83  & 4.34  & -     & -     & -     & -     & -     & -     & 12.58 & 11.91 & 12.05 & 10.62 & 10.24 & 11.14 \\
          &       & Cln\{$0^\circ,270^\circ$\} & 4.97  & 4.63  & 5.01  & 4.27  & 4.64  & 4.29  & -     & -     & -     & -     & -     & -     & 12.46 & 13.88 & 10.62 & 9.67  & 9.89  & 9.88 \\
   \rowcolor{gray}        &       & Cln\{$180^\circ,90^\circ$\} & 6.15  & 6.01  & 5.95  & 4.32  & 4.83  & 4.34  & -     & -     & -     & -     & -     & -     & 21.08 & 21.57 & 16.84 & 14.30 & 16.53 & 13.75 \\
          &       & Frank & 7.84  & 8.03  & 7.64  & 3.33  & 3.23  & 3.59  & -     & -     & -     & -     & -     & -     & 12.82 & 11.16 & 12.88 & 11.71 & 11.56 & 12.00 \\
    \bottomrule
    \end{tabular}
   \end{small}  
  \begin{flushleft}
\begin{footnotesize}
Cln\{$\omega_1^\circ,\omega_2^\circ$\}: The bivariate copulas are the Clayton  rotated by $\omega_1^\circ$ and $\omega_2^\circ$ to handle the positive and negative dependencies, respectively.
\end{footnotesize}  
\end{flushleft}    
\end{table}
\end{landscape}

\begin{landscape}
\begin{table}[htbp]
  \centering
  \caption{\label{sim2}Small sample of sizes $N = 50$ simulations ($10^4$ replications, $n_q=25$) from the one-factor copula mixed model  with beta margins and biases,  root mean square errors (RMSEs) and standard deviations (SDs), along with the square root of the average theoretical variances ($\sqrt{\bar V}$), scaled by 100, for the MLEs   under different copula choices and margins.  The true (simulated) copula distributions are the Clayton  and the Clayton copula rotated by $270^\circ$ to handle the positive and negative dependencies, respectively. }
\begin{small}

    \begin{tabular}{lllcccccccccccccccccc}
    \toprule
   {} & margin & copula & $\pi_{11}$ & $\pi_{12}$ & $\pi_{13}$ & $\pi_{01}$ & $\pi_{02}$ & $\pi_{03}$ & $\g_{11}$ & $\g_{12}$ & $\g_{13}$ & $\g_{01}$ & $\g_{02}$ & $\g_{03}$ & $\tau_{11}$ & $\tau_{12}$ & $\tau_{13}$ & $\tau_{01}$ & $\tau_{02}$ & $\tau_{03}$ \\
    \midrule
 \rowcolor{gray}    Bias  & normal & BVN   & 2.93  & 1.97  & 2.95  & 2.24  & 3.25  & 2.24  & -     & -     & -     & -     & -     & -     & 1.84  & -0.68 & 2.45  & -1.61 & -1.47 & -1.42 \\
          &       & Cln\{$0^\circ,270^\circ$\} & 3.06  & 2.14  & 3.08  & 2.26  & 3.21  & 2.26  & -     & -     & -     & -     & -     & -     & 0.91  & 0.64  & 0.94  & 0.20  & -0.05 & 0.45 \\
 \rowcolor{gray}          &       & Cln\{$180^\circ,90^\circ$\} & 2.35  & 1.15  & 2.53  & 2.44  & 3.57  & 2.31  & -     & -     & -     & -     & -     & -     & 1.68  & 1.12  & -2.09 & 5.04  & 6.03  & 2.97 \\
          &       & Frank & 0.59  & -0.94 & 0.80  & 3.80  & 4.70  & 3.33  & -     & -     & -     & -     & -     & -     & 3.79  & 1.06  & 4.32  & -2.75 & -2.95 & -2.17 \\
     \rowcolor{gray}      & beta  & BVN   & 0.02  & -0.04 & 0.01  & -0.05 & -0.05 & -0.02 & -0.51 & -0.39 & -0.44 & -0.12 & 0.05  & -0.18 & 3.82  & 1.22  & 4.27  & -2.42 & -2.15 & -2.13 \\
          &       & Cln\{$0^\circ,270^\circ$\} & 0.02  & -0.04 & 0.00  & -0.02 & 0.01  & 0.00  & -0.32 & -0.31 & -0.30 & -0.26 & -0.25 & -0.25 & 2.68  & 2.78  & 1.93  & -0.46 & -0.93 & -0.21 \\
  \rowcolor{gray}         &       & Cln\{$180^\circ,90^\circ$\} & -0.92 & -1.13 & -0.76 & 0.06  & 0.02  & 0.02  & 0.71  & 0.98  & 0.65  & 0.53  & 1.01  & 0.22  & 2.44  & 2.70  & -1.50 & 5.52  & 6.99  & 3.61 \\
          &       & Frank & -2.27 & -2.72 & -2.10 & 1.45  & 1.52  & 1.00  & 0.33  & 0.32  & 0.28  & -0.07 & 0.01  & -0.17 & 5.71  & 3.13  & 6.05  & -3.57 & -4.10 & -2.71 \\
 \rowcolor{gray}    SD    & normal & BVN   & 2.06  & 2.47  & 2.05  & 2.40  & 2.00  & 2.40  & 11.66 & 9.89  & 11.74 & 10.04 & 12.32 & 9.89  & 18.72 & 20.68 & 16.77 & 14.64 & 16.23 & 13.67 \\
          &       & Cln\{$0^\circ,270^\circ$\} & 2.07  & 2.48  & 2.07  & 2.41  & 2.00  & 2.42  & 11.27 & 9.61  & 11.38 & 10.02 & 12.09 & 9.87  & 18.46 & 20.82 & 15.71 & 13.45 & 15.73 & 12.19 \\
 \rowcolor{gray}          &       & Cln\{$180^\circ,90^\circ$\} & 2.26  & 2.71  & 2.21  & 2.49  & 2.11  & 2.46  & 14.95 & 13.24 & 14.47 & 11.67 & 15.20 & 10.84 & 26.52 & 28.18 & 22.62 & 15.73 & 17.69 & 14.99 \\
          &       & Frank & 2.75  & 3.23  & 2.72  & 2.81  & 2.32  & 2.72  & 12.88 & 10.95 & 12.78 & 10.41 & 12.88 & 10.12 & 19.16 & 20.63 & 17.70 & 15.43 & 16.95 & 14.40 \\
 \rowcolor{gray}          & beta  & BVN   & 2.00  & 2.27  & 1.98  & 2.20  & 1.92  & 2.21  & 2.18  & 2.11  & 2.19  & 2.08  & 2.23  & 2.05  & 12.50 & 13.02 & 11.69 & 11.38 & 11.23 & 11.74 \\
          &       & Cln\{$0^\circ,270^\circ$\} & 2.00  & 2.28  & 1.99  & 2.20  & 1.94  & 2.22  & 2.13  & 2.03  & 2.15  & 2.00  & 2.08  & 2.01  & 13.11 & 14.29 & 11.50 & 10.26 & 10.78 & 10.40 \\
  \rowcolor{gray}         &       & Cln\{$180^\circ,90^\circ$\} & 2.34  & 2.59  & 2.28  & 2.23  & 1.96  & 2.24  & 2.95  & 2.96  & 2.86  & 2.42  & 2.77  & 2.25  & 23.13 & 23.40 & 19.66 & 13.78 & 14.60 & 13.98 \\
          &       & Frank & 2.79  & 3.03  & 2.76  & 2.49  & 2.13  & 2.44  & 2.62  & 2.47  & 2.57  & 2.11  & 2.22  & 2.07  & 12.09 & 11.55 & 12.06 & 12.18 & 12.00 & 12.52 \\
  \rowcolor{gray}   $\sqrt{\bar V}$  & normal & BVN   & 1.90  & 2.35  & 1.91  & 2.33  & 1.91  & 2.32  & 11.06 & 9.40  & 11.21 & 9.25  & 11.25 & 9.17  & 11.32 & 12.80 & 10.33 & 10.02 & 9.80  & 10.39 \\
          &       & Cln\{$0^\circ,270^\circ$\} & 1.86  & 2.29  & 1.87  & 2.29  & 1.86  & 2.30  & 10.27 & 8.76  & 10.52 & 9.06  & 10.92 & 9.06  & 11.78 & 14.25 & 10.16 & 9.46  & 9.69  & 9.43 \\
      \rowcolor{gray}     &       & Cln\{$180^\circ,90^\circ$\} & 2.08  & 2.58  & 2.06  & 2.40  & 1.97  & 2.38  & 12.64 & 10.91 & 12.49 & 10.08 & 12.43 & 9.80  & 14.47 & 15.07 & 13.34 & 10.86 & 11.33 & 11.36 \\
          &       & Frank & 2.10  & 2.46  & 2.11  & 2.29  & 1.79  & 2.32  & 11.73 & 9.86  & 11.83 & 9.39  & 11.48 & 9.24  & 10.36 & 10.72 & 10.17 & 10.42 & 9.89  & 10.95 \\
  \rowcolor{gray}         & beta  & BVN   & 1.89  & 2.19  & 1.90  & 2.16  & 1.90  & 2.15  & 2.08  & 2.04  & 2.11  & 1.99  & 2.10  & 1.97  & 11.14 & 12.61 & 10.15 & 9.97  & 9.70  & 10.36 \\
          &       & Cln\{$0^\circ,270^\circ$\} & 1.82  & 2.10  & 1.83  & 2.11  & 1.83  & 2.12  & 1.93  & 1.87  & 1.97  & 1.89  & 1.94  & 1.91  & 12.19 & 14.62 & 10.45 & 9.28  & 9.53  & 9.27 \\
  \rowcolor{gray}         &       & Cln\{$180^\circ,90^\circ$\} & 2.12  & 2.42  & 2.10  & 2.22  & 1.96  & 2.20  & 2.52  & 2.51  & 2.49  & 2.24  & 2.45  & 2.16  & 14.54 & 15.22 & 13.29 & 11.06 & 11.55 & 11.60 \\
          &       & Frank & 2.05  & 2.26  & 2.07  & 2.10  & 1.76  & 2.14  & 2.30  & 2.20  & 2.31  & 1.98  & 2.06  & 1.96  & 10.35 & 10.78 & 10.09 & 10.44 & 9.94  & 10.97 \\
  \rowcolor{gray}   RMSE  & normal & BVN   & 3.58  & 3.16  & 3.59  & 3.28  & 3.81  & 3.28  & -     & -     & -     & -     & -     & -     & 18.81 & 20.70 & 16.95 & 14.73 & 16.30 & 13.74 \\
          &       & Cln\{$0^\circ,270^\circ$\} & 3.69  & 3.28  & 3.71  & 3.30  & 3.78  & 3.31  & -     & -     & -     & -     & -     & -     & 18.48 & 20.83 & 15.73 & 13.45 & 15.73 & 12.20 \\
   \rowcolor{gray}        &       & Cln\{$180^\circ,90^\circ$\} & 3.27  & 2.94  & 3.36  & 3.49  & 4.15  & 3.38  & -     & -     & -     & -     & -     & -     & 26.58 & 28.20 & 22.71 & 16.52 & 18.69 & 15.28 \\
          &       & Frank & 2.81  & 3.36  & 2.83  & 4.72  & 5.24  & 4.30  & -     & -     & -     & -     & -     & -     & 19.53 & 20.66 & 18.22 & 15.67 & 17.21 & 14.56 \\
   \rowcolor{gray}        & beta  & BVN   & 2.00  & 2.27  & 1.98  & 2.20  & 1.92  & 2.21  & 2.24  & 2.15  & 2.24  & 2.08  & 2.23  & 2.06  & 13.07 & 13.07 & 12.45 & 11.63 & 11.44 & 11.93 \\
          &       & Cln\{$0^\circ,270^\circ$\} & 2.00  & 2.28  & 1.99  & 2.20  & 1.94  & 2.22  & 2.16  & 2.05  & 2.17  & 2.01  & 2.10  & 2.03  & 13.38 & 14.56 & 11.66 & 10.27 & 10.82 & 10.40 \\
  \rowcolor{gray}         &       & Cln\{$180^\circ,90^\circ$\} & 2.51  & 2.83  & 2.40  & 2.23  & 1.96  & 2.24  & 3.04  & 3.12  & 2.93  & 2.48  & 2.95  & 2.26  & 23.26 & 23.56 & 19.71 & 14.85 & 16.19 & 14.44 \\
          &       & Frank & 3.60  & 4.07  & 3.47  & 2.88  & 2.61  & 2.64  & 2.64  & 2.49  & 2.59  & 2.11  & 2.22  & 2.08  & 13.37 & 11.96 & 13.49 & 12.69 & 12.68 & 12.81 \\
    \bottomrule
    \end{tabular}
    \end{small}
  \begin{flushleft}
\begin{footnotesize}
Cln\{$\omega_1^\circ,\omega_2^\circ$\}: The bivariate copulas are the Clayton  rotated by $\omega_1^\circ$ and $\omega_2^\circ$ to handle the positive and negative dependencies, respectively.
\end{footnotesize}  
\end{flushleft}    
\end{table}
\end{landscape}

\begin{itemize}

\item The MLEs of $\bfpi_1,\bfpi_0$  are rather robust to bivariate copula misspecification,  but their biases  increase when the assumed bivariate copulas  have different  tail dependence behaviour. For example, in Table \ref{sim1}  (Table \ref{sim2}) the  scaled biases for the MLEs of $\pi_{11}$ for the various one-factor copula mixed models  with normal (beta) margins increase to $-1.38$ (-0.92) and $-3.71$ (-2.27) when rotated Clayton copulas with opposite direction tail dependence and Frank copulas with tail independence, respectively, are called. 

\item  The MLEs  of $\debf_1,\debf_0$  are rather robust to bivariate copula misspecification,  but their biases  increase when the assumed bivariate copula  has tail dependence of opposite direction from the true bivariate copula. For example, in  Table \ref{sim1}  (Table \ref{sim2})  the  scaled biases for the MLEs of $\sigma_{02}$ ($\gamma_{02}$) for the various one-factor copula mixed models  with normal (beta) margins range from $-0.83$ ( $-0.25$ ) to $1.75$ ($0.05$), but the scaled bias
 increases to $6.85$ ($1.01$) when rotated Clayton copulas with opposite direction tail dependence are called.

\item The  ML  estimates of $\tau$'s   are robust to  margin misspecification, as the copula remains invariant under any series of strictly increasing transformations of the components of the random vector, e.g.,  in Table \ref{sim1} the scaled bias of $\hat\tau_{13}$ is $1.81$ for the true  one-factor copula mixed model and $1.57$  for an one-factor  copula mixed model with the true bivariate copulas but  beta margins.

 \end{itemize}

\subsection{\label{miss}Misspecification  of the copula-mixed model -- Sensitivity analysis to the conditional independence }
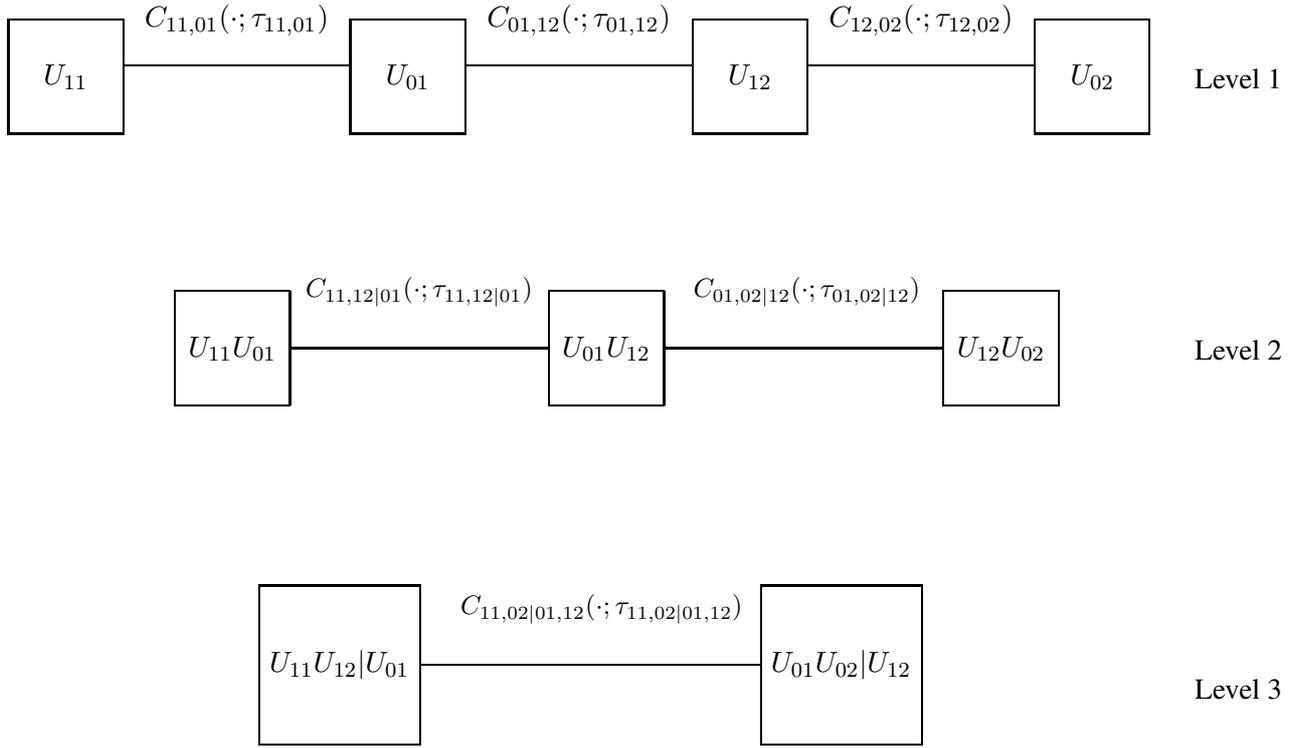
\begin{figure}[!h]
\vspace{1cm}
\begin{center}
\setlength{\unitlength}{3.cm}
\begin{picture}(5,1)
\put(-0.3,0.7){\framebox(0.5,0.5){$U_{11}$}}
\put(0.2,1){\line(1,0){1}}
\put(0.7,1.2){\makebox(0,0){$C_{11,01}(\cdot;\tau_{11,01})$}}
\put(1.2,0.7){\framebox(0.5,0.5){$U_{01}$}}
\put(1.7,1){\line(1,0){1}}
\put(2.2,1.2){\makebox(0,0){$C_{01,12}(\cdot;\tau_{01,12})$}}
\put(2.7,0.7){\framebox(0.5,0.5){$U_{12}$}}
\put(3.2,1){\line(1,0){1}}
\put(3.7,1.2){\makebox(0,0){$C_{12,02}(\cdot;\tau_{12,02})$}}
\put(4.2,0.7){\framebox(0.5,0.5){$U_{02}$}}

\put(4.9,0.9){Level 1}

\put(0.43,-0.5){\framebox(0.5,0.5){$U_{11}U_{01}$}}
\put(0.93,-0.25){\line(1,0){1.14}}
\put(1.5,-0.){\makebox(0,0){\small$C_{11,12|01}(\cdot;\tau_{11,12|01})$}}
\put(2.07,-0.5){\framebox(0.5,0.5){$U_{01}U_{12}$}}
\put(2.58,-0.25){\line(1,0){1.21}}
\put(3.2,0.){\makebox(0,0){\small$C_{01,02|12}(\cdot;\tau_{01,02|12})$}}
\put(3.8,-0.5){\framebox(0.5,0.5){$U_{12}U_{02}$}}

\put(4.9,-0.3){Level 2}

\put(0.8,-2){\framebox(0.7,0.7){$U_{11}U_{12}|U_{01}$}}
\put(1.5,-1.65){\line(1,0){1.5}}
\put(2.3,-1.4){\makebox(0,0){\small$C_{11,02|01,12}(\cdot;\tau_{11,02|01,12})$}}
\put(3.,-2){\framebox(0.7,0.7){$U_{01}U_{02}|U_{12}$}}

\put(4.9,-1.8){Level 3}

\end{picture}
\end{center}
\vspace{6cm}
\caption{\label{4dvine} Graphical representation of the 4-dimensional D-vine copula model with 3 levels.}
\end{figure}

We show a sensitivity analysis to the conditional independence assumption. 
We randomly generate 10,000 samples   from the D-vine copula mixed model with  both  normal (Table \ref{sim3}) and beta (Table \ref{sim4}) margins using the algorithm in   \cite{nikoloulopoulos-2018-smmr}. 
We set the sample size $N$, the study size $n$, the true univariate and Kendall's $\tau$  parameters, and the disease prevalence  to mimic the rheumatoid arthritis data in \cite{Nishimura-etal2007}. The D-vine copula mixed model assumes full dependence among the tests as the D-vine copula is not truncated, i.e.,  there are  bivariate copulas  not only at level 1 of the D-vine. Figure \ref{4dvine} depicts the representation of the D-vine copula model. The copulas at the higher levels model the conditional dependence. 
The true (simulated) D-vine copula mixed  model uses  Clayton copulas rotated by 270$^\circ$ at level 1 and Clayton copulas at levels 2 and 3.

We have estimated  the one-factor copula mixed model with different bivariate copulas and margins. 
Tables \ref{sim3} and  \ref{sim4} contain the resultant biases, root mean square errors (RMSEs) and standard deviations (SDs), along with average theoretical variances, scaled by 100, for the MLEs of the common parameters  under different copula choices and margins. The theoretical variances of the MLEs are obtained via the gradients and the Hessian that were computed numerically during the maximization process.  

\setlength{\tabcolsep}{8pt}
\begin{table}[!h]
  \centering
  \caption{\label{sim3}Small sample of sizes $N = 22$ simulations ($10^4$ replications, $n_q=25$) from the D-vine copula  mixed model   with normal margins and biases,  root mean square errors (RMSEs) and standard deviations (SDs), along with the square root of the average theoretical variances ($\sqrt{\bar V}$), scaled by 100, for the MLEs  of the one-factor copula mixed model  under different copula choices and margins.  The true (simulated) D-vine copula mixed  model uses  Clayton copulas rotated by 270$^\circ$ at level 1 and Clayton copulas at levels 2 and 3.}
  \begin{small}

    \begin{tabular}{lllcccccccc}
    \toprule
          & margin & copula & $\pi_{11}$ & $\pi_{12}$ & $\pi_{01}$ & $\pi_{02}$ & $\s_{11}$ & $\s_{12}$ & $\s_{01}$ & $\s_{02}$ \\
    \midrule
  \rowcolor{gray}   Bias  & normal & BVN   & -0.07 & -0.29 & -0.21 & -0.05 & -3.29 & -4.01 & -2.07 & -1.70 \\
          &       & Cln\{$0^\circ,270^\circ$\} & -0.10 & -0.29 & -0.32 & -0.06 & -2.76 & -3.36 & -1.11 & -2.07 \\
 \rowcolor{gray}          &       & Cln\{$180^\circ,90^\circ$\} & -0.68 & -0.87 & -0.10 & -0.03 & -0.15 & -1.04 & -0.84 & 0.14 \\
          &       & Frank & -1.33 & -1.47 & 0.06  & 0.03  & -2.53 & -3.21 & -1.69 & -1.69 \\
  \rowcolor{gray}         & beta  & BVN   & -1.58 & -1.71 & -4.31 & -1.22 & -     & -     & -     & - \\
           &       & Cln\{$0^\circ,270^\circ$\} & -1.64 & -1.75 & -4.46 & -1.24 & -     & -     & -     & - \\
  \rowcolor{gray}         &       & Cln\{$180^\circ,270^\circ$\} & -2.39 & -2.50 & -4.08 & -1.16 & -     & -     & -     & - \\
          &       & Frank & -2.64 & -2.73 & -4.14 & -1.15 & -     & -     & -     & - \\
  \rowcolor{gray}   SD    & normal & BVN   & 3.61  & 3.48  & 3.39  & 0.84  & 12.80 & 12.27 & 18.92 & 19.00 \\
          &       & Cln\{$0^\circ,270^\circ$\} & 3.62  & 3.50  & 3.42  & 0.84  & 12.84 & 12.35 & 19.79 & 19.23 \\
 \rowcolor{gray}          &       & Cln\{$180^\circ,90^\circ$\} & 3.89  & 3.76  & 3.49  & 0.86  & 15.16 & 14.69 & 19.73 & 20.00 \\
           &       & Frank & 4.06  & 3.93  & 3.58  & 0.88  & 13.43 & 12.89 & 19.19 & 19.18 \\
  \rowcolor{gray}         & beta  & BVN   & 3.35  & 3.25  & 3.58  & 1.06  & 2.68  & 2.55  & 4.02  & 1.51 \\
          &       & Cln\{$0^\circ,270^\circ$\} & 3.37  & 3.28  & 3.65  & 1.10  & 2.76  & 2.63  & 4.26  & 1.57 \\
  \rowcolor{gray}         &       & Cln\{$180^\circ,90^\circ$\} & 3.72  & 3.62  & 3.61  & 1.09  & 3.38  & 3.23  & 4.11  & 1.62 \\
         &       & Frank & 3.74  & 3.67  & 3.83  & 1.13  & 2.89  & 2.76  & 4.20  & 1.55 \\
   \rowcolor{gray}  $\sqrt{\bar V}$  & normal & BVN   & 3.38  & 3.31  & 3.22  & 0.80  & 12.07 & 11.84 & 17.15 & 17.57 \\
          &       & Cln\{$0^\circ,270^\circ$\} & 3.22  & 3.15  & 3.04  & 0.77  & 10.75 & 10.51 & 15.55 & 16.50 \\
    \rowcolor{gray}       &       & Cln\{$180^\circ,90^\circ$\} & 3.26  & 3.19  & 3.15  & 0.81  & 11.78 & 11.55 & 16.34 & 16.85 \\
         &       & Frank & 3.34  & 3.25  & 3.11  & 0.78  & 11.95 & 11.67 & 16.55 & 17.34 \\
     \rowcolor{gray}      & beta  & BVN   & 3.14  & 3.08  & 3.17  & 0.94  & 2.64  & 2.56  & 3.48  & 1.26 \\
           &       & Cln\{$0^\circ,270^\circ$\} & 2.97  & 2.91  & 2.89  & 0.88  & 2.32  & 2.25  & 3.04  & 1.12 \\
   \rowcolor{gray}        &       & Cln\{$180^\circ,90^\circ$\} & 3.04  & 2.98  & 3.06  & 0.95  & 2.57  & 2.47  & 3.31  & 1.25 \\
          &       & Frank & 3.07  & 3.01  & 3.01  & 0.90  & 2.63  & 2.55  & 3.27  & 1.19 \\
  \rowcolor{gray}   RMSE  & normal & BVN   & 3.61  & 3.49  & 3.40  & 0.84  & 13.22 & 12.91 & 19.03 & 19.07 \\
          &       & Cln\{$0^\circ,270^\circ$\} & 3.62  & 3.51  & 3.43  & 0.85  & 13.14 & 12.80 & 19.82 & 19.34 \\
 \rowcolor{gray}          &       & Cln\{$180^\circ,90^\circ$\} & 3.95  & 3.86  & 3.49  & 0.86  & 15.16 & 14.73 & 19.75 & 20.00 \\
          &       & Frank & 4.27  & 4.19  & 3.58  & 0.88  & 13.67 & 13.28 & 19.27 & 19.25 \\
   \rowcolor{gray}        & beta  & BVN   & 3.71  & 3.67  & 5.60  & 1.62  & -     & -     & -     & - \\
          &       & Cln\{$0^\circ,270^\circ$\} & 3.74  & 3.71  & 5.76  & 1.65  & -     & -     & -     & - \\
  \rowcolor{gray}         &       & Cln\{$180^\circ,90^\circ$\} & 4.43  & 4.40  & 5.59  & 1.64  & -     & -     & -     & - \\          &       & Frank & 4.58  & 4.57  & 5.64  & 1.61  & -     & -     & -     & - \\
    \bottomrule
    \end{tabular}%
    \end{small}
  \begin{flushleft}
\begin{footnotesize}
Cln\{$\omega_1^\circ,\omega_2^\circ$\}: The bivariate copulas are the Clayton  rotated by $\omega_1^\circ$ and $\omega_2^\circ$ to handle the positive and negative dependencies, respectively.
\end{footnotesize}  
\end{flushleft}    
\end{table}

\begin{table}[!h]
  \centering
  \caption{\label{sim4}Small sample of sizes $N = 22$ simulations ($10^4$ replications, $n_q=25$) from the D-vine copula  mixed model   with beta margins and biases,  root mean square errors (RMSEs) and standard deviations (SDs), along with the square root of the average theoretical variances ($\sqrt{\bar V}$), scaled by 100, for the MLEs  of the one-factor copula mixed model  under different copula choices and margins.  The true (simulated) D-vine copula mixed  model uses  Clayton copulas rotated by 270$^\circ$ at level 1 and Clayton copulas at levels 2 and 3. }
  \begin{small}

    \begin{tabular}{lllcccccccc}
    \toprule
          & margin & copula & $\pi_{11}$ & $\pi_{12}$ & $\pi_{01}$ & $\pi_{02}$ & $\g_{11}$ & $\g_{12}$ & $\g_{01}$ & $\g_{02}$ \\
    \midrule
 \rowcolor{gray}    Bias  & normal & BVN   & 1.95  & 1.74  & 5.08  & 0.78  & -     & -     & -     & - \\
          &       & Cln\{$0^\circ,270^\circ$\} & 2.00  & 1.79  & 4.99  & 0.78  & -     & -     & -     & - \\
  \rowcolor{gray}         &       & Cln\{$180^\circ,90^\circ$\} & 1.43  & 1.23  & 5.17  & 0.80  & -     & -     & -     & - \\
          &       & Frank & 0.53  & 0.34  & 5.29  & 0.84  & -     & -     & -     & - \\
  \rowcolor{gray}         & beta  & BVN   & 0.11  & -0.05 & 0.03  & -0.02 & -0.67 & -0.88 & -0.63 & -0.06 \\
          &       & Cln\{$0^\circ,270^\circ$\} & 0.13  & -0.03 & -0.06 & -0.02 & -0.66 & -0.81 & -0.47 & -0.07 \\
    \rowcolor{gray}       &       & Cln\{$180^\circ,90^\circ$\} & -0.68 & -0.83 & 0.08  & -0.01 & 0.32  & 0.10  & -0.42 & 0.03 \\
          &       & Frank & -1.18 & -1.33 & 0.25  & 0.07  & -0.36 & -0.57 & -0.52 & -0.06 \\
 \rowcolor{gray}    SD    & normal & BVN   & 3.69  & 3.64  & 3.18  & 0.69  & 14.01 & 13.64 & 24.77 & 19.13 \\
          &       & Cln\{$0^\circ,270^\circ$\} & 3.70  & 3.65  & 3.23  & 0.69  & 13.85 & 13.56 & 25.90 & 18.99 \\
   \rowcolor{gray}        &       & Cln\{$180^\circ,90^\circ$\} & 3.96  & 3.91  & 3.27  & 0.71  & 16.74 & 16.51 & 25.56 & 20.12 \\
          &       & Frank & 4.17  & 4.15  & 3.31  & 0.72  & 14.60 & 14.19 & 25.01 & 19.02 \\
  \rowcolor{gray}         & beta  & BVN   & 3.38  & 3.35  & 3.27  & 0.74  & 2.84  & 2.76  & 4.36  & 0.95 \\
          &       & Cln\{$0^\circ,270^\circ$\} & 3.37  & 3.33  & 3.32  & 0.75  & 2.83  & 2.77  & 4.52  & 0.96 \\
    \rowcolor{gray}       &       & Cln\{$180^\circ,90^\circ$\} & 3.71  & 3.69  & 3.31  & 0.75  & 3.62  & 3.54  & 4.51  & 1.04 \\
          &       & Frank & 3.83  & 3.80  & 3.42  & 0.76  & 3.08  & 2.98  & 4.46  & 0.95 \\
    \rowcolor{gray}$\sqrt{\bar V}$  & normal & BVN   & 3.48  & 3.46  & 2.88  & 0.64  & 12.92 & 12.87 & 22.18 & 18.45 \\
          &       & Cln\{$0^\circ,270^\circ$\} & 3.28  & 3.25  & 2.67  & 0.62  & 11.41 & 11.29 & 20.07 & 18.01 \\
   \rowcolor{gray}        &       & Cln\{$180^\circ,90^\circ$\} & 3.39  & 3.37  & 2.79  & 0.64  & 12.92 & 12.73 & 21.20 & 18.13 \\
          &       & Frank & 3.45  & 3.41  & 2.77  & 0.62  & 12.77 & 12.67 & 21.54 & 18.55 \\
 \rowcolor{gray}          & beta  & BVN   & 3.21  & 3.19  & 3.07  & 0.72  & 2.82  & 2.78  & 4.04  & 0.97 \\
          &       & Cln\{$0^\circ,270^\circ$\} & 3.00  & 2.98  & 2.74  & 0.70  & 2.43  & 2.40  & 3.52  & 0.91 \\
    \rowcolor{gray}       &       & Cln\{$180^\circ,90^\circ$\} & 3.13  & 3.12  & 2.99  & 0.73  & 2.78  & 2.76  & 3.96  & 0.99 \\
            &       & Frank & 3.15  & 3.13  & 2.86  & 0.69  & 2.80  & 2.76  & 3.77  & 0.93 \\
   \rowcolor{gray}  RMSE  & normal & BVN   & 4.17  & 4.03  & 5.99  & 1.04  & -     & -     & -     & - \\
          &       & Cln\{$0^\circ,270^\circ$\} & 4.21  & 4.07  & 5.94  & 1.04  & -     & -     & -     & - \\
   \rowcolor{gray}        &       & Cln\{$180^\circ,90^\circ$\} & 4.21  & 4.10  & 6.12  & 1.07  & -     & -     & -     & - \\
          &       & Frank & 4.20  & 4.16  & 6.24  & 1.11  & -     & -     & -     & - \\
 \rowcolor{gray}          & beta  & BVN   & 3.38  & 3.35  & 3.27  & 0.74  & 2.92  & 2.90  & 4.40  & 0.96 \\
          &       & Cln\{$0^\circ,270^\circ$\} & 3.37  & 3.33  & 3.32  & 0.75  & 2.90  & 2.88  & 4.54  & 0.96 \\
  \rowcolor{gray}         &       & Cln\{$180^\circ,90^\circ$\} & 3.77  & 3.78  & 3.32  & 0.75  & 3.64  & 3.54  & 4.53  & 1.04 \\
 
          &       & Frank & 4.01  & 4.03  & 3.43  & 0.76  & 3.11  & 3.04  & 4.49  & 0.96 \\
    \bottomrule
    \end{tabular}%
     \end{small}
 \begin{flushleft}
\begin{footnotesize}
Cln\{$\omega_1^\circ,\omega_2^\circ$\}: The bivariate copulas are the Clayton  rotated by $\omega_1^\circ$ and $\omega_2^\circ$ to handle the positive and negative dependencies, respectively.
\end{footnotesize}  
\end{flushleft}    
\end{table}

From Table \ref{sim3} (Table \ref{sim4}) it is seen that the one-factor copula mixed model with normal (beta) margins led to unbiased  and efficient estimates when 
the bivariate copulas are  a combination of Clayton  and rotated Clayton by $270^\circ$ to model the  positive and negative dependencies, respectively. These are the same with the true (simulated) copulas of the D-vine copula mixed model which imply that the sensitivity and specificity of each test have tail dependence.   Hence, the tail dependence between the factor and each of the latent sensitivities/specificities is inherited to the tail dependence between the latent sensitivities and specificities, and thus, the conditional independence assumption has no impact on the estimation of the meta-analytic parameters of sensitivity and specificity of each test when this assumption is violated.  This is due the fact that the one-factor copula can be explained as an 1-truncated C-vine rooted at the factor \citep{krupskii-joe-2013,nikoloulopoulos&joe12,Kadhem&Nikoloulopoulos-2021}. 
Note also that in line with the results in the preceding subsection, the biases of the estimates increase when the assumed bivariate copulas  have  tail dependence of opposite direction from the true copulas or tail independence. When the BVN copulas with intermediate tail dependence are used to link the factor with the latent sensitivities/specificities, the estimates are robust to misspecification of the copula mixed  model as long as the univariate margins are correctly specified.

Finally in order to study the relative performance of the one-factor copula mixed model over the quadrivariate vine copula mixed model  as the number of quadrature points  increase we randomly generated $B = 20$ samples of size $N = 22$ from the D-vine copula mixed model.  The  model parameters  are set as before. The simulations were carried out on a 
Broadwell E5-2680 v4@2.40GHz.
Table \ref{times} summarizes the computing times (averaged over 20 replications) in seconds. Clearly the D-vine copula mixed approach requires a much higher computing time.   
Hence it is demonstrated that even for the case of $T=2$ tests, the computational improvement of the one-factor copula mixed model is substantial, as one has to calculate numerically bivariate integrals instead of  much more difficult quadrivariate integrals.

\begin{table}[!h]
  \centering
  \caption{\label{times}Small sample of size $N=22$ simulations (20 replications) from the quadrivariate D-vine copula mixed model
 and computing times (averaged over 20 replications) in seconds of the one-factor and quadrivariate D-vine copula mixed approaches. }
\begin{small}   
\setlength{\tabcolsep}{70pt}
    \begin{tabular}{ccc}
    \toprule
    $n_q$ & factor & vine \\
    \midrule
    15    & 35.4  & 799.9 \\
    30    & 65.7  & 7355.4 \\
    50    & 126.2 & 42997.6 \\
    \bottomrule
    \end{tabular}
    \end{small} 
\end{table}

\section{\label{app-sec}  Application }
\cite{Nishimura-etal2007} contacted a systematic review and summarized data of rheumatoid factor (RF) and anti-cyclic citrullinated peptide (anti-CCP)  antibodies for diagnosing rheumatoid arthritis.  They included $N=22$  studies that assessed both RF and anti-CCP2 antibody for diagnosing rheumatoid arthritis 
and  used the 1987 revised American College of Rheumatology (ACR) criteria as the perfect reference standard of rheumatoid arthritis \citep{Arnett-etal-1988}. These data have been frequently used as an example for methodological papers on joint meta-analysis of diagnostic accuracy studies in a multiple tests design with a gold standard (e.g., \citealt{dimou-etal2016,nikoloulopoulos-2018-smmr}).  \cite{liu-etal-2015-biometrics} in one of their examples  deal with the same  
data, but as they propose models for the meta-analysis of the accuracy of a diagnostic  test under evaluation and an imperfect  reference test, they use only the  the RF test as the  index test for detection of rheumatoid arthritis and assume that  the ACR 1987 revised criteria are an imperfect reference   test for classification.
Their analysis  confirmed that the ACR 1987 revised criteria are a prefect reference test as the estimates of sensitivity and specificity of the ACR 1987 criteria (reference test) were 1, suggesting that such reference test is in fact a gold standard.

We use  the one-factor copula mixed model in order to determine whether anti-CCP antibody  identifies more accurately patients with rheumatoid arthritis than RF does.
We fit the one-factor copula mixed model for all  choices of parametric families of copulas and margins. To make it easier
to compare strengths of dependence, we convert from the BVN, Frank  and   (rotated)  Clayton 
$\hat\theta$'s to $\hat\tau$'s  via the relations in (\ref{tauBVN}), (\ref{tauFrank}), and (\ref{tauCln}). Because the number of parameters is the same between the models, we use the log-likelihood at the maximum likelihood estimates as a rough diagnostic measure for model selection between the models. 
For vine copulas (one-factor copula is an 1-truncated C-vine copula), \cite{Dissmann-etal-2013-csda} found that pair-copula selection based on likelihood  seems to be better than even using bivariate goodness-of-fit tests.  The goodness-of-fit procedures involve a global distance measure between the model-based and empirical distribution, hence they might not be sensitive to tail behaviours and are not diagnostic in the sense of suggesting improved parametric models in the case of small $p$-values \citep[page 254]{joe2014}.  A larger  likelihood  value indicates a  model that better approximates both  the dependence structure of the data and the strength of dependence in the tails.

\begin{table}[!h]
  \centering
  \caption{\label{arthritis-res}Maximized log-likelihoods, estimates and standard errors (SE) of the one-factor copula mixed models for the rheumatoid arthritis data.}
\begin{small}

\setlength{\tabcolsep}{5.5pt}
    \begin{tabular}{lccccccccccccccc}
    \toprule
          &       & \multicolumn{2}{c}{BVN} &       & \multicolumn{2}{c}{Frank} &       & \multicolumn{2}{c}{Cln\{$0^\circ,90^\circ$\}} &       & \multicolumn{2}{c}{Cln\{$0^\circ,270^\circ$\}} &       & \multicolumn{2}{c}{Cln\{$180^\circ,270^\circ$\}} \\
\cmidrule{3-4}  \cmidrule{6-7} \cmidrule{9-10}   \cmidrule{12-13}  \cmidrule{15-16}  
          &       & Est.  & SE    &       & Est.  & SE    &       & Est.  & SE    &       & Est.  & SE    &       & Est.  & SE \\  \midrule
   \multicolumn{16}{l}{Normal margins} \\
    $\pi_{11}$ &       & 0.681 & 0.034 &       & 0.660 & 0.033 &       & 0.678 & 0.033 &       & 0.681 & 0.036 &       & 0.676 & 0.034 \\
    $\pi_{12}$ &       & 0.684 & 0.034 &       & 0.655 & 0.031 &       & 0.673 & 0.032 &       & 0.675 & 0.034 &       & 0.674 & 0.035 \\
    $\pi_{01}$ &       & 0.825 & 0.033 &       & 0.834 & 0.032 &       & 0.827 & 0.033 &       & 0.826 & 0.033 &       & 0.827 & 0.033 \\
    $\pi_{02}$ &       & 0.960 & 0.008 &       & 0.962 & 0.000 &       & 0.960 & 0.008 &       & 0.960 & 0.008 &       & 0.960 & 0.008 \\
    $\s_{11}$ &       & 0.685 & 0.128 &       & 0.698 & 0.134 &       & 0.691 & 0.122 &       & 0.722 & 0.133 &       & 0.687 & 0.129 \\
    $\s_{12}$ &       & 0.697 & 0.124 &       & 0.675 & 0.123 &       & 0.657 & 0.112 &       & 0.687 & 0.121 &       & 0.722 & 0.134 \\
    $\s_{01}$ &       & 1.028 & 0.181 &       & 1.028 & 0.177 &       & 1.037 & 0.183 &       & 1.029 & 0.181 &       & 1.027 & 0.178 \\
    $\s_{02}$ &       & 0.790 & 0.175 &       & 0.795 & 0.164 &       & 0.794 & 0.184 &       & 0.792 & 0.170 &       & 0.797 & 0.171 \\
    $\tau_{11}$ &       & 0.644 & 0.168 &       & 0.680 & 0.119 &       & 0.719 & 0.137 &       & 0.716 & 0.124 &       & 0.818 & 0.223 \\
    $\tau_{12}$ &       & 0.802 & 0.395 &       & 0.839 & 0.152 &       & 0.750 & 0.149 &       & 0.826 & 0.144 &       & 0.466 & 0.136 \\
    $\tau_{01}$ &       & -0.125 & 0.168 &       & -0.218 & 0.160 &       & -0.149 & 0.161 &       & -0.213 & 0.148 &       & -0.227 & 0.162 \\
    $\tau_{02}$ &       & -0.201 & 0.182 &       & -0.289 & 0.183 &       & -0.228 & 0.333 &       & -0.272 & 0.203 &       & -0.278 & 0.221 \\
    $-\log(L)$ &       & \multicolumn{2}{c}{322.4} &       & \multicolumn{2}{c}{321.0} &       & \multicolumn{2}{c}{320.1} &       & \multicolumn{2}{c}{318.9} &       & \multicolumn{2}{c}{325.3} \\
    Beta margins &       &       &       &       &       &       &       &       &       &       &       &       &       &       &  \\
    $\pi_{11}$ &       & 0.667 & 0.031 &       & 0.648 & 0.032 &       & 0.664 & 0.033 &       & 0.665 & 0.031 &       & 0.661 & 0.032 \\
    $\pi_{12}$ &       & 0.670 & 0.032 &       & 0.646 & 0.032 &       & 0.661 & 0.032 &       & 0.661 & 0.030 &       & 0.658 & 0.033 \\
    $\pi_{01}$ &       & 0.782 & 0.034 &       & 0.789 & 0.033 &       & 0.783 & 0.034 &       & 0.784 & 0.033 &       & 0.785 & 0.033 \\
    $\pi_{02}$ &       & 0.949 & 0.009 &       & 0.950 & 0.009 &       & 0.949 & 0.009 &       & 0.949 & 0.009 &       & 0.949 & 0.009 \\
    $\s_{11}$ &       & 0.087 & 0.028 &       & 0.092 & 0.030 &       & 0.089 & 0.030 &       & 0.097 & 0.029 &       & 0.089 & 0.029 \\
    $\s_{12}$ &       & 0.091 & 0.028 &       & 0.092 & 0.027 &       & 0.083 & 0.028 &       & 0.091 & 0.026 &       & 0.098 & 0.032 \\
    $\s_{01}$ &       & 0.132 & 0.039 &       & 0.132 & 0.039 &       & 0.133 & 0.039 &       & 0.132 & 0.039 &       & 0.130 & 0.039 \\
    $\s_{02}$ &       & 0.025 & 0.012 &       & 0.026 & 0.013 &       & 0.025 & 0.012 &       & 0.026 & 0.013 &       & 0.027 & 0.013 \\
    $\tau_{11}$ &       & 0.635 & 0.226 &       & 0.937 & 0.004 &       & 0.723 & 0.140 &       & 0.731 & 0.128 &       & 0.815 & 0.231 \\
    $\tau_{12}$ &       & 0.849 & 0.644 &       & 0.651 & 0.103 &       & 0.764 & 0.168 &       & 0.811 & 0.126 &       & 0.497 & 0.134 \\
    $\tau_{01}$ &       & -0.111 & 0.169 &       & -0.175 & 0.167 &       & -0.120 & 0.164 &       & -0.217 & 0.144 &       & -0.234 & 0.173 \\
    $\tau_{02}$ &       & -0.203 & 0.179 &       & -0.195 & 0.187 &       & -0.212 & 0.290 &       & -0.248 & 0.192 &       & -0.278 & 0.221 \\
    $-\log(L)$ &       & \multicolumn{2}{c}{323.3} &       & \multicolumn{2}{c}{322.8} &       & \multicolumn{2}{c}{321.2} &       & \multicolumn{2}{c}{320.1} &       & \multicolumn{2}{c}{326.3} \\
    \bottomrule
    \end{tabular}
    \end{small} 
\begin{flushleft}
\begin{footnotesize}
Cln\{$\omega_1^\circ,\omega_2^\circ$\}: The bivariate copulas are the Clayton  rotated by $\omega_1^\circ$ and $\omega_2^\circ$ to handle the positive and negative dependencies, respectively.
\end{footnotesize}  
\end{flushleft}    
\end{table}

The log-likelihoods showed that an one-factor copula mixed model with Clayton and Clayton rotated by 270$^\circ$ degrees copulas with normal margins to join the factor with each of the sensitivities/specificities  provides the best fit (Table \ref{arthritis-res}). For this particular example it is revealed that an one-factor copula mixed model with the sensitivities and specificities on the transformed  scale provides better fit  than an one-factor copula mixed model with beta margins, which models the sensitivity and specificity on the original  scale.

The resultant sensitivities and  specificities show that the anti-CCP2 antibody is better compared with RF. Both tests have fairly similar sensitivity but the anti-CCP2 is much more specific.  On the one hand, the estimated univariate parameters and standard errors are in line with the ones in  \cite{nikoloulopoulos-2018-smmr}, but the implementation of the proposed model  is much faster, since a numerically time-consuming four-dimensional integral calculation is replaced with a numerically  fast  two-dimensional integral calculation on the other.

\begin{figure}[!h]
\begin{center}
\begin{footnotesize}
\begin{tabular}{|cc|}
\hline
Rheumatoid Factor & Anti-CCP2 antibody\\\hline

\includegraphics[width=0.45\textwidth]{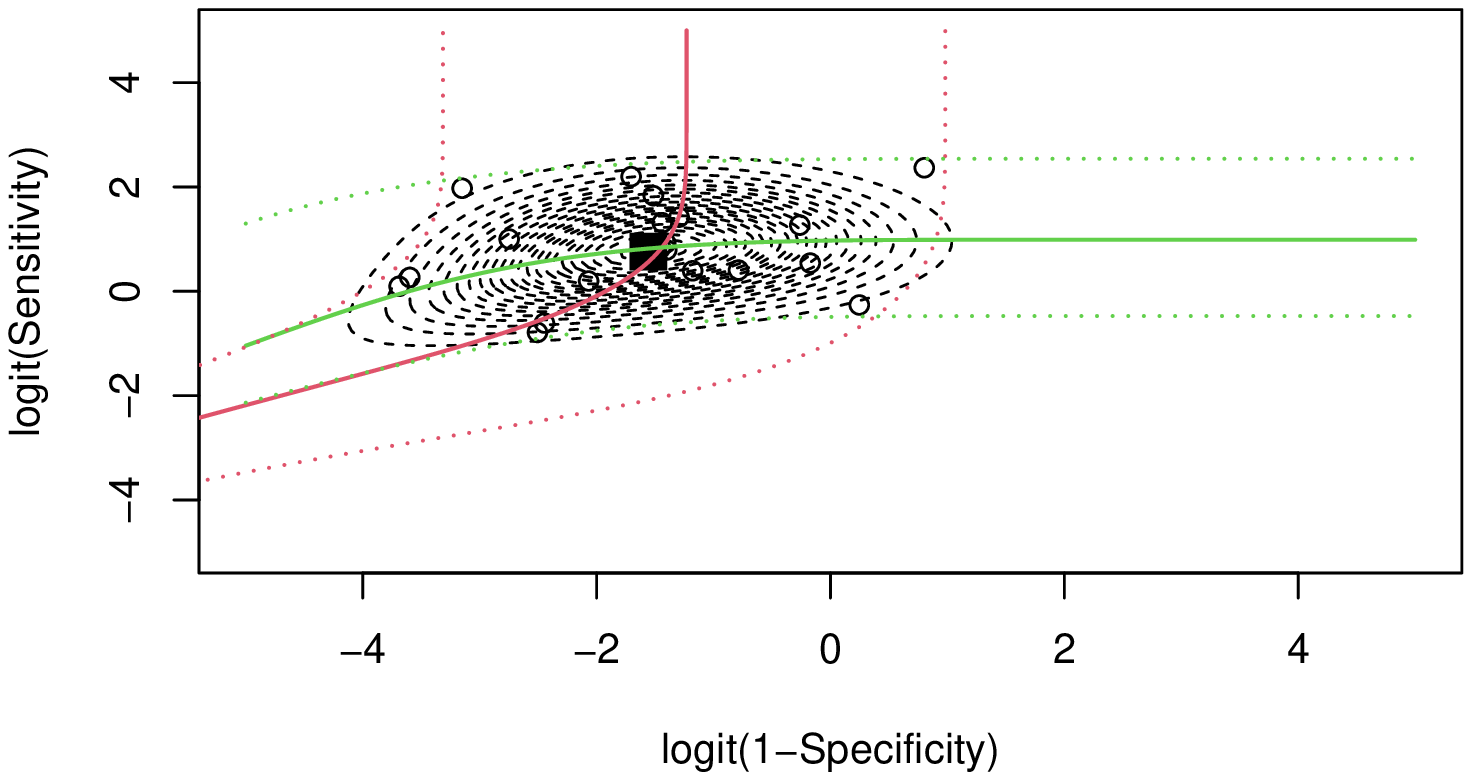}
&

\includegraphics[width=0.45\textwidth]{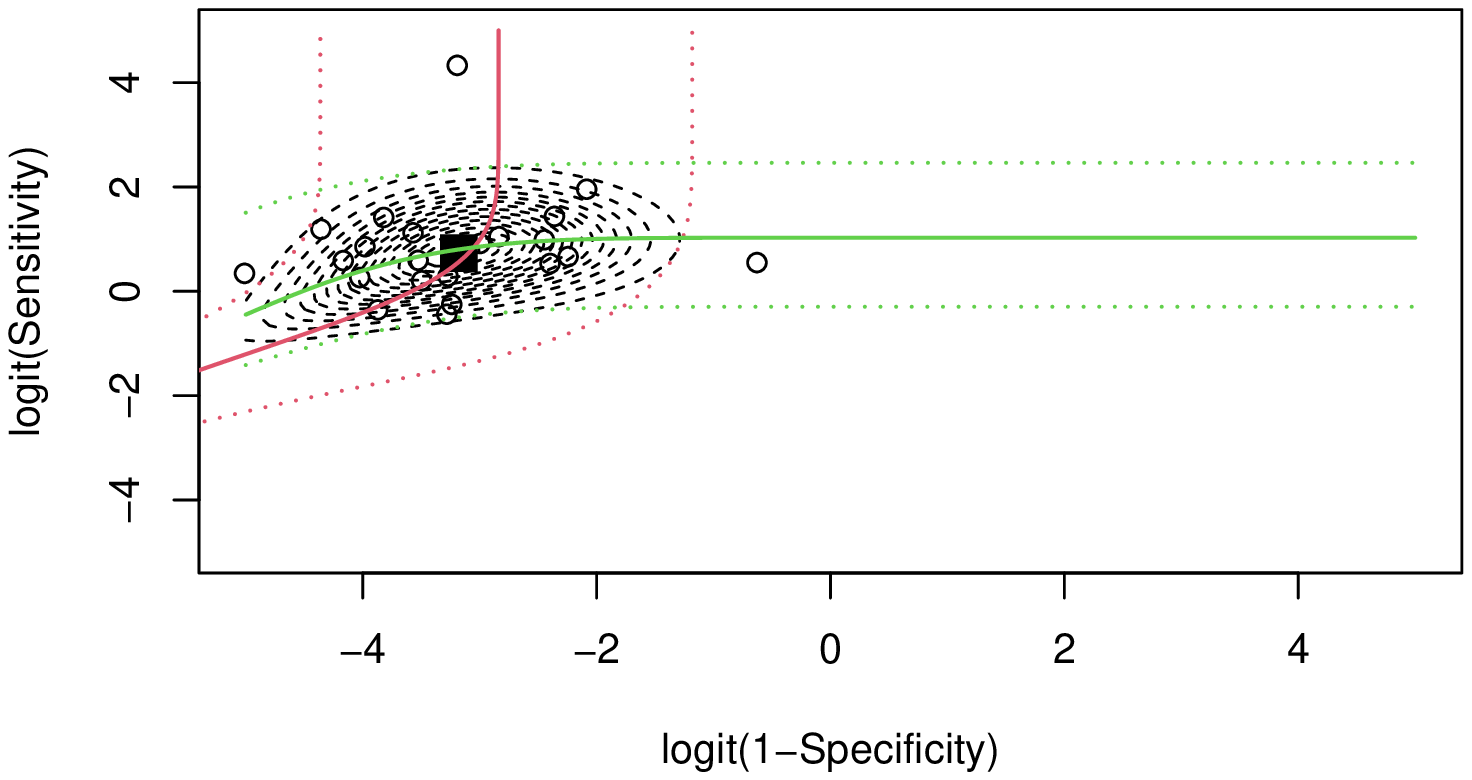}\\\hline
\end{tabular}
\end{footnotesize}
\caption{\label{ROC-arthritis}Contour plots (predictive region)  and quantile  regression curves  from the best fitted one-factor copula mixed model for the  rheumatoid arthritis data. Red and green lines represent the quantile  regression curves $x_{1t}:=\widetilde{x}_{1t}(x_{0t},q)$ and $x_{0t}:=\widetilde{x}_{0t}(x_{1t},q)$, respectively; for $q=0.5$ solid lines and for $q\in\{0.01,0.99\}$ dotted lines (confidence region). The axes are in  logit scale since  we also plot  the estimated contour plot of the random effects distribution as predictive region; this has been estimated for the logit pair of (Sensitivity, Specificity) for each test.}
\end{center}

\end{figure}

\begin{figure}[!h]
\begin{center}

\begin{tabular}{|cc|}
\hline
$x_{1t}:=\widetilde{x}_{1t}(x_{0t},q=0.5)$ & $x_{0t}:=\widetilde{x}_{0t}(x_{1t},q=0.5)$\\\hline

\includegraphics[width=0.45\textwidth]{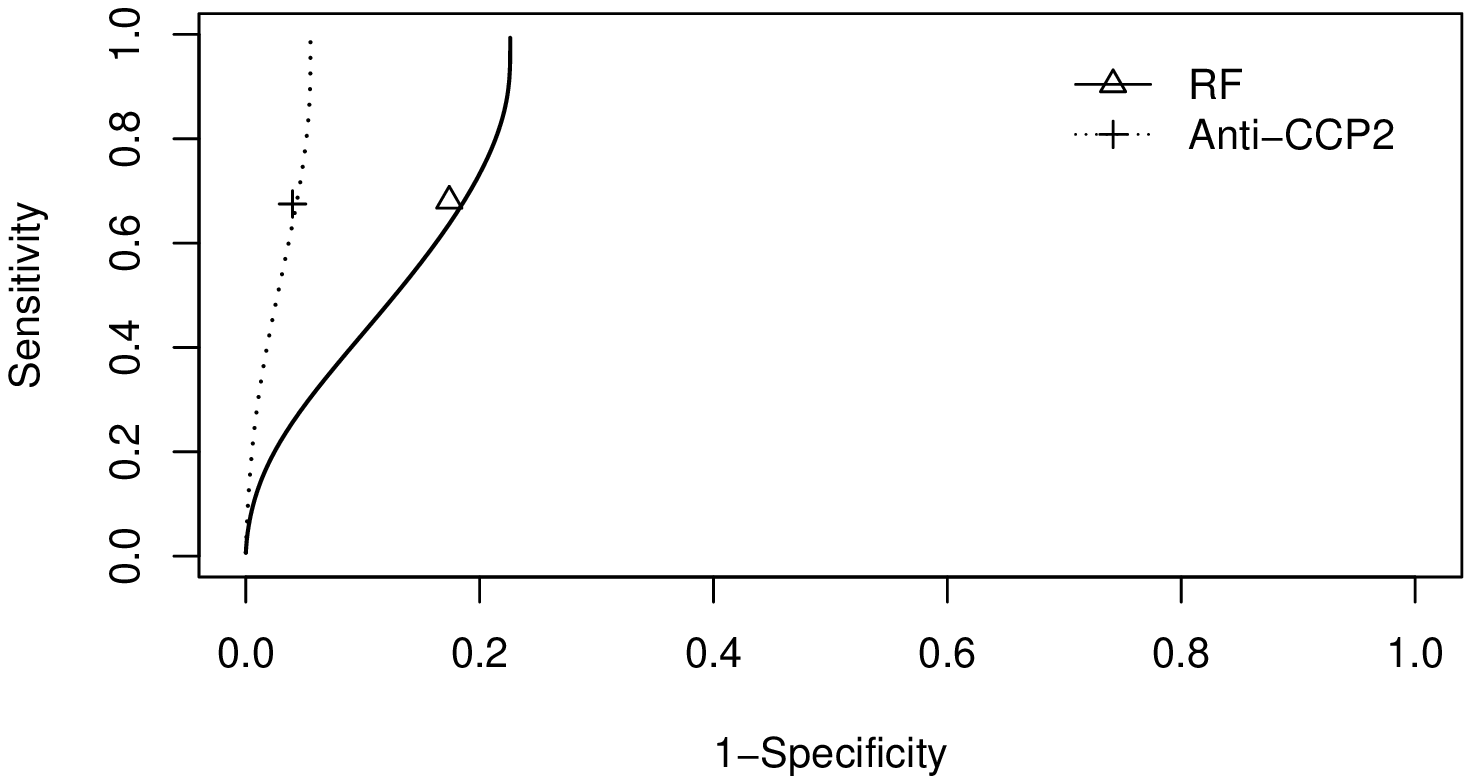}
&

\includegraphics[width=0.45\textwidth]{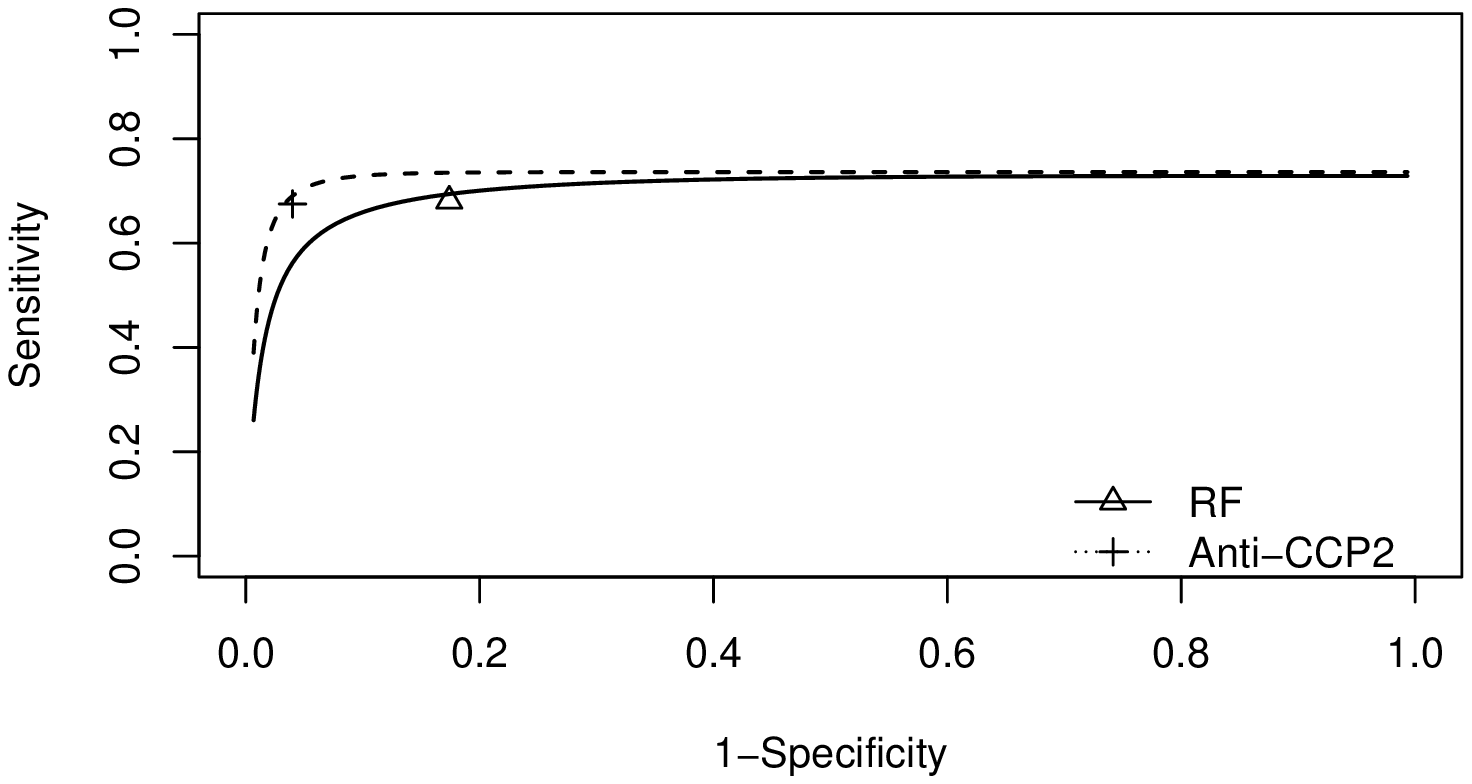}\\\hline
\end{tabular}

\caption{\label{ROC-comp-arthritis}Median regression curves for each test backtransformed to the original scale of sensitivity and specificity for the rheumatoid arthritis data.}
\end{center}

\end{figure}

From the Kendall's tau estimates and standard errors there is strong evidence of  dependence between the two diagnostic tests.  The fact that the best-fitting bivariate copulas are Clayton and  Clayton rotated by 270$^\circ$ reveals that there is  tail dependence  among the latent sensitivities and specificities. 
This can be further seen trough the predictive region of  the SROC curves. 
Figure \ref{ROC-arthritis} depicts the  SROC curves and summary operating points (a pair of average sensitivity and specificity) with a confidence region and a predictive region for each test from the best fitted one-factor copula mixed model.  
Sharper corners in the predictive region indicate tail dependence. 
Figure \ref{ROC-comp-arthritis} provides a direct and visual comparison between the two competing diagnostic tests and reveals that the anti-CCP2 antibody is better compared with RF.

\section{\label{discussion}Discussion}

\baselineskip=23pt

We have proposed an one-factor copula mixed model for joint meta-analysis and comparison of  multiple diagnostic tests in a multiple tests design with a gold standard. This is a parsimonious meta-analytic model that (a)  has the $2T$-variate GLMM with an additive latent structure as a special case when the BVN copulas are used, (b) can have a latent structure that is not additive if other than BVN copulas are called, (c) can model the latent sensitivities and specificities on the original scale rather than a transformed scale as in the $2T$-variate GLMM (d) enables the meta-analytic parameters of interest to be separated from the copula (dependence) parameters which are interpretable as dependence of the latent sensitivity/specificity with another latent variable, (e) avoids  the curse of multi-dimensionality and (f) models  adequately  the dependence among the latent sensitivities and specificities as it can be explained as an 1-truncated C-vine copula.

Our model  
can 
provide an improvement over the $2T$-variate GLMM  with an additive latent structure as  the random effects distribution is expressed via an one-factor copula that provides  a wide range of dependence with $2T$ dependence parameters and allow for different types of tail behaviour, different from assuming simple linear correlation structures, normality and tail independence. This strength of multivariate meta-analysis  approaches that use copulas has been  pointed out by 
\cite{Jacson&White-2018-BiomJ} and \cite{jackson&white&riley-2020} and it has also  been exploited 
 in network  meta-analysis 
 \citep{Phillippo-etal-2020-jrssa}.

The $2T$-variate D-vine copula mixed model, which it has as special case the $2T$-variate GLMM with an unstructured correlation structure,  provides full dependence, but it is intractable as the number of competing tests increases. The $2T$-variate one-factor copula mixed model solves this problem since
the joint likelihood reduces to an one-dimensional integral of a function which in turn is a product of $2T$ one-dimensional integrals, hence the method avoids $2T$-dimensional integration which is time consuming even for $T=2$ tests. 
Its parsimony  is not a distributional concern about the dependence between the tests 
due to the main result in \cite{joeetal10}:   all the bivariate margins of the vine copula 
have (tail) dependence if the bivariate copulas at level 1 have (tail) dependence. This is satisfied by the one-factor copula as it is an 1-truncated C-vine. 
Hence, the proposed model  can form the vehicle  for conducting meta-analysis of  comparative accuracy studies with  three or more tests.

When the focus is on estimates of the meta-analytic univariate parameters of interest, the outgrowth of joint analysis is modest, in that the differences in the summary estimates and standard errors from separate meta-analyses for each test are not that distinct. The most striking differences between separate and joint meta-analyses arise when one deduces comparative diagnostic accuracy, i.e., an SROC curve.  
 An SROC curve makes much more sense and will help decision makers to assess the actual diagnostic accuracy of the competing diagnostic tests. In an era of evidence-based medicine, decision makers need high-quality procedures such as the SROC curves to support decisions about whether or not to use a diagnostic test in a specific clinical situation and, if so, which test.
We have deduced  SROC curves from the  one-factor copula mixed model. The model parameters (including dependence parameters), the choice of the copula, and the choice of the margin affect the shape of the SROC curves.  
A series of independence  models cannot be used to produce the SROC curves, since the dependence parameters affect the shape of the SROC curve and these are set to independence.

Comparative accuracy studies with paired designs where each test is applied to the same patients  should report the  data as separate $2\times 2$ tables. 
 Authors of primary studies of diagnostic accuracy that assess three or more tests in the same patients should be encouraged to report sufficient data to extract separate $2\times 2$ tables of test results as in Table \ref{2times2T}.  Comparative accuracy studies should rightly use  a multiple tests  designs  so that patients receive each test in order to reduce biases and ensure the clinical relevance of the resulting inferences \citep{trikalinos-etal-2014-rsm}. 
 
Nevertheless, in practice there exist comparative studies in a randomized  design  or even non-comparative studies  \citep{Takwoingi-etal-2013}  and for some of them  the reference test might be imperfect. Future research will focus on  extending the one-factor copula mixed model  to incorporate    randomised designs and non-comparative studies  with or without a gold standard. \cite{Ma-etal-2018-biostatistics} and \cite{Lian-eta-al-2019-jasa}     proposed methods for comparing multiple diagnostic tests that can   incorporate studies with different designs and studies with our without gold standard. As their methods assume that the between-studies model is the multivariate normal distribution that suffers for the curse of multidimensionality when the numbers of tests increases, we will exploit the use of the one-factor copula distribution.
The one-factor copula distribution  will  provide computational  and distributional improvements 
when adopted to the setting of \cite{Ma-etal-2018-biostatistics} and \cite{Lian-eta-al-2019-jasa}.

\section*{Software}

{\tt R} functions to implement the one-factor copula mixed model for meta-analysis of multiple diagnostic tests will be part of the next major release of the  {\tt  R} package {\tt  CopulaREMADA}  \citep{Nikoloulopoulos-2018-CopulaREMADA}.

\section*{Acknowledgements}

The simulations presented in this paper were carried out on the High Performance Computing Cluster supported by the Research and Specialist Computing Support service at the University of East Anglia.

\baselineskip=20pt


\end{document}